%% file: ms.tex
\newif\ifSingleColumn
\SingleColumnfalse

\ifSingleColumn
\documentclass[draftclsnofoot, onecolumn, 12pt]{IEEEtran}
\else
\documentclass[twocolumn,journal,10pt]{IEEEtran}
\fi
\usepackage[draft]{hyperref}

\input{header.tex}

\input{defines.tex}

\begin{document}

\title{{Cross-Domain Continual Learning for \\Edge Intelligence in Wireless ISAC Networks}} 

\ifodd 1
\author{
\IEEEauthorblockN{
\normalsize{Jingzhi~Hu},~\IEEEmembership{\normalsize Member,~IEEE},
\normalsize{Xin~Li},~\IEEEmembership{\normalsize Member,~IEEE},
\normalsize{Zhou Su},~\IEEEmembership{\normalsize Senior~Member,~IEEE},
and~\normalsize{Jun~Luo},~\IEEEmembership{\normalsize Fellow,~IEEE}
}
\thanks{This work has been submitted to the IEEE for possible publication. Copyright may be transferred without notice, after which this version may no longer be accessible.}
\thanks{
J. Hu, X. Li, and J. Luo are with the College of Computing and Data Science, Nanyang Technological University, Singapore 639798~(email: jingzhi.hu518@gmail.com, l.xin@ntu.edu.sg, junluo@ntu.edu.sg).
 }
\thanks{Z. Su is with the School of Cyber Science and Engineering, Xi'an Jiaotong University, Xi'an 710049, China (e-mail: zhousu@ieee.org).}
}
\else
\author{\linespread{1.25}
\IEEEauthorblockN{
\normalsize{Submission under preparation for IEEE TWC.}
}
}
\fi
\maketitle
\begin{abstract}
\input{./sections/1_abstract.tex}

\end{abstract}

\input{sections/2_introduction.tex}

\input{sections/3_sysmodel.tex}

\input{sections/4_problem.tex}

\input{sections/5_algorithm.tex}
\input{sections/6_simulation.tex}

\input{sections/7_conclusion.tex}

\begin{appendices}
\input{sections/appendices/app_prop.tex}
\end{appendices}

\balance
\renewcommand{\refname}{References} 
\bibliographystyle{IEEEtran}
\bibliography{bibilio}

\end{document}

%% file: header.tex
\usepackage[utf8]{inputenc}
\usepackage{glossaries}
\usepackage[symbols,nogroupskip]{glossaries-extra}
\usepackage{bm}
\usepackage{amssymb}
\usepackage{amsmath}
\usepackage{amsfonts}
\usepackage{mathrsfs}
\usepackage{bbm}
\usepackage{scalerel} 
\usepackage{graphicx}
\usepackage{subfigure}
\usepackage{cite}
\usepackage{enumerate}
\usepackage{url}
\usepackage{epstopdf}
\usepackage{algorithm}
\usepackage{algpseudocode} 
\usepackage{algorithmicx,algorithm}
\usepackage{float}
\usepackage{subfigure}
\usepackage{amsthm}
\usepackage{color}
\usepackage{enumitem}
\usepackage{clipboard}
\usepackage[normalem]{ulem} 
\usepackage{framed}
\usepackage{array} 
\usepackage{hhline}
\usepackage[table]{xcolor}
\usepackage{balance}
\usepackage{xspace}
\usepackage{dblfloatfix} 

\usepackage{xargs} 
\usepackage{cooltooltips}
\usepackage{xcolor}  
\usepackage{comment}
\usepackage{caption}

\newcommand{\beq}{\begin{equation}}
\newcommand{\eeq}{\end{equation}}
\newcommand{\beqq}{\begin{equation*}}
\newcommand{\eeqq}{\end{equation*}}
\newcommand{\mbeqa}{\begin{align}}
\newcommand{\meeqa}{\end{align}}

\makeatletter
\def\@eqnnum{{\normalfont \normalcolor \theequation}}
\makeatother

\makeatletter
\newcommand{\manuallabel}[2]{\def\@currentlabel{#2}\label{#1}}
\makeatother

\makeatletter
\newcommand{\customlabel}[2]{%
   \protected@write \@auxout {}{\string \newlabel {#1}{{#2}{\thepage}{#2}{#1}{}} }%
   \hypertarget{#1}{}
}
\makeatother

\DeclareMathOperator*{\argmin}{argmin}
\DeclareMathOperator*{\argmax}{argmax}

\newtheorem{proposition}{\bf Proposition}

\newtheorem{corollary}{\bf Corollary}

\allowdisplaybreaks

\newcommand{\frev}{\textcolor[rgb]{0,0,0}}

%% file: defines.tex
\newcommand{\tp}{{\scriptscriptstyle\top}}
\newcommand{\iu}{\mathrm{i}}

\renewcommand{\Pr}{\operatorname{Pr}}

\newcommand{\name}{EdgeCL\xspace}

\newcommand{\norm}{\mathtt{norm}}

\newcommand{\rT}{\mathrm{T}}

\newcommand{\cD}{\mathcal{D}}
\newcommand{\cC}{\mathcal{C}}
\newcommand{\cL}{\mathcal{L}}
\newcommand{\cR}{\mathcal{R}}
\newcommand{\cF}{\mathcal{F}}
\newcommand{\cS}{\mathcal{S}}
\newcommand{\cK}{\mathcal{K}}
\newcommand{\cO}{\mathcal{O}}

\newcommand{\bfv}{\bm f}

\newcommand{\bgun}{\mathbf{g}}

\newcommand{\btheta}{\bm\theta}
\newcommand{\bdelta}{\bm\delta}

\newcommand{\rTx}{\mathrm{Tx}} 
\newcommand{\rRx}{\mathrm{Rx}} 
\newcommand{\rstatic}{\mathrm{S}} 
\newcommand{\rdyna}{\mathrm{D}} 
\newcommand{\rU}{\mathrm{U}}
\newcommand{\rH}{\mathrm{H}}

\newcommand{\rP}{\mathrm{P}}

\newcommand{\rin}{\mathrm{in}}
\newcommand{\rout}{\mathrm{out}}
\newcommand{\rQ}{\scriptscriptstyle\mathrm{Q}} 
\newcommand{\rK}{\scriptscriptstyle\mathrm{K}} 
\newcommand{\rV}{\scriptscriptstyle\mathrm{V}} 
\newcommand{\bfg}{\mathbf{g}}
\newcommand{\tAttn}{\mathtt{Attn}}
\newcommand{\csim}{\!\sim\!}



%% file: sections/1_abstract.tex
In wireless networks with integrated sensing and communications (ISAC), edge intelligence~(EI) is expected to be developed at edge devices~(ED) for sensing user activities based on channel state information~(CSI).
However, due to the CSI being highly specific to users' characteristics, the CSI-activity relationship is notoriously domain dependent, essentially demanding EI to learn sufficient datasets from various domains in order to gain cross-domain sensing capability.
This poses a crucial challenge owing to the EDs' limited resources, for which storing datasets across all domains will be a significant burden.
In this paper, we propose the \emph{\name} framework, enabling the EI to continually learn-then-discard each incoming dataset, while remaining resilient to catastrophic forgetting.
We design a transformer-based discriminator for handling sequences of noisy and nonequispaced CSI samples.
Besides, we propose a distilled core-set based knowledge retention method with robustness-enhanced optimization to train the discriminator, preserving its performance for previous domains while preventing future forgetting.
Experimental evaluations show that \name achieves 89\% of performance compared to cumulative training while consuming only 3\% of its memory, mitigating forgetting by 79\%.

%% file: sections/2_introduction.tex
\section[Introduction]{Introduction}
\label{sec: intro}

Integrated sensing and communications~(ISAC) techniques are widely recognized as a crucial enabler for 6G communications~\cite{Dong23TWC_Sensing} 
as it helps establish cognitive and perceptive wireless networks in a cost-efficient manner without requiring extensive additional infrastructure~\cite{Liu22JSAC_ISAC}.
Specifically, by utilizing the readily available channel state information (CSI), ISAC enables edge devices (EDs), such as cellular base stations and Wi-Fi access points, to perceive channel variations caused by human motions and environment dynamics.
By handling the channel variations with powerful signal processing techniques empowered by edge intelligence~(EI), i.e., the artificial intelligence on the edge, the EDs can develop various sensing functionalities.
These sensing functionalities, in which human activity recognition~(HAR) is a prominent example~\cite{Li23COMMAG_Integrated}, can retrospectively benefit communications through improved context and mobility awareness~\cite{Strinati24EuCNC_Distributed,Saleem21TWC_Mobility}.
With advances in EI and the increasingly large bandwidth and antenna arrays used by wireless transceivers, the sensing capability of EDs in wireless ISAC networks is expected to grow rapidly, paving the way for more intelligent and adaptive wireless networks.

Nevertheless, the EI tends to be highly \emph{domain-dependent}.
\Copy{E-1-1}{\frev{According to~\cite{Akrout23CST_Domain}, a domain in this context can be defined as the statistical relationship between CSI variations and user activities.
In practice, the domains should be designated as the variable that changes during the EI's deployment and whose change has a significant influence on this CSI-activity relationship.}}
Existing studies pursue the cross-domain sensing capability of EI through extracting domain-invariant features from CSI, either by using the Doppler frequency shifts~(DFS)~\cite{Zhang21TPAMI_Widar3,Niu22TMC_Understand} or adversarial learning~\cite{Jiang18MobiCom_Sulu, Wang23TMC_AirFi, Li21IMWUT_CrossGR}.
However, domain-invariant features discard domain-relevant information in CSI~\cite{Bui21NIPS_Exploiting} and thus suffer from deficient accuracy.
Moreover, in common situations where users are in close proximity to their devices, their individual characteristics, such as body contours and subtle movements, have a significant impact on CSI~\cite{Hu2023Mobicom_Muse}, which can hardly be removed without substantially degrading the sensing potential of CSI.
Due to the unpredictable nature of environment dynamics and user characteristics in practice, substantial discrepancies exist between the domains in offline pre-training and those encountered in online deployments, making it challenging for EI to obtain robust and high-accuracy cross-domain sensing capabilities.

Essentially, to guarantee practical cross-domain sensing capability, the EI needs to actually learn for the domains encountered during online deployment~\cite{Gulrajani21ICLR_In}.
This requires the collection of sufficient training datasets and timely update of the EI to extend its sensing capability to new domains. 
However, as new domains continually emerge, the CSI datasets expand over time; storing these datasets in an ED, which generally has rather limited memory resources, becomes increasingly prohibitive~\cite{Mao17CST_Survey,Haibeh22Access_Survey}, especially when CSI data is collected over large bandwidths of multiple bands for high sensing precision~\cite{Li24MobiSys_UWB}.
Although it is possible for the ED to upload the datasets and its EI to a cloud server and conduct training remotely to alleviate its local burden~\cite{Lin19PIEEE_Computation}, this approach will inevitably incur substantial communication overheads to the backbone network and lead to high latency~\cite{Barbera13INFOCOM_Offload}.

\begin{figure}[t] 
    \centering
    \includegraphics[width=0.9\linewidth]{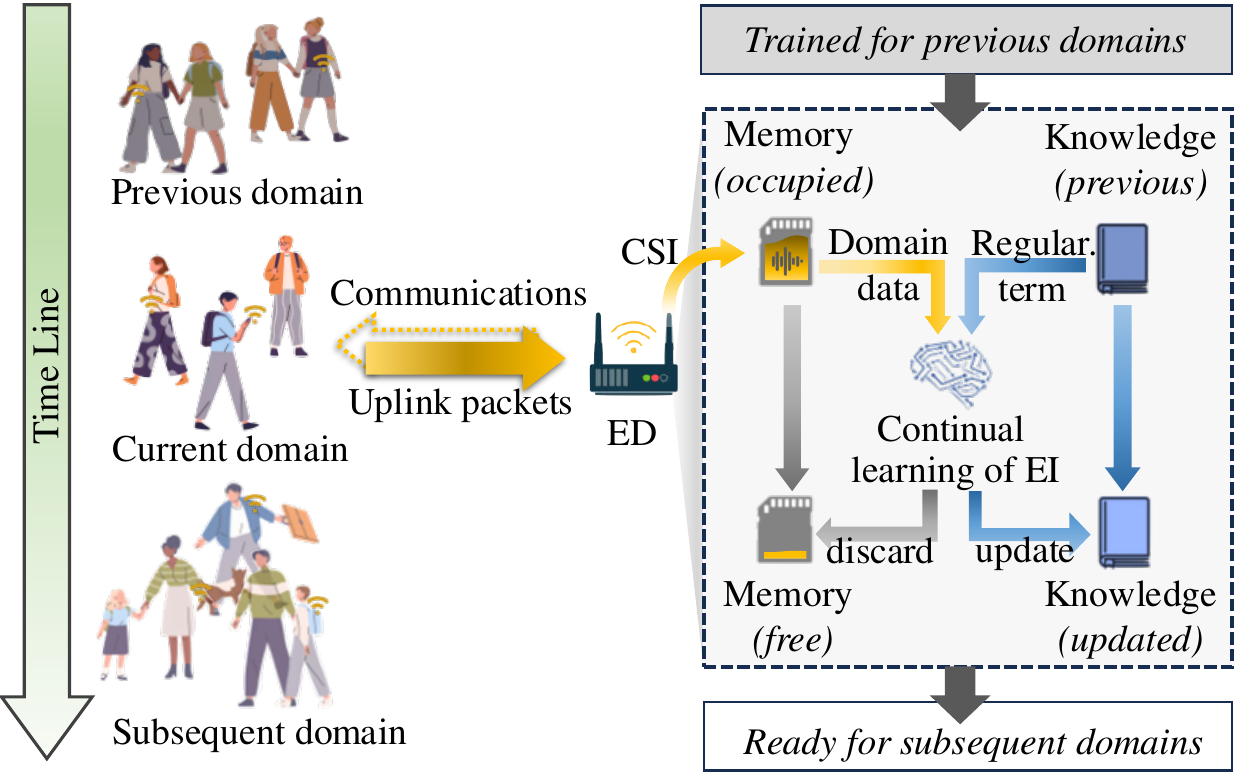}
        \caption{Working principle of \name: The CSI dataset of current domain and the knowledge learned from previous domains are jointly used for the cross-domain continual learning of EI.}
        \label{fig_sys_mod}
    \vspace{-.5em}
    \end{figure}

Fortunately, as different domains emerge at different times, the datasets for each domain, referred to as \emph{domain datasets}, are gathered sequentially. 
This allows an ED to train its EI on each domain dataset one by one.
For example, consider a scenario where a new domain dataset is collected periodically to accommodate newly emerged users within the period.
At the end of each period, the EI can be trained solely on this new domain dataset, which is then discarded to free up memory resources on the ED, preparing for subsequent domain datasets.
Nevertheless, such sequential training suffers from the \emph{catastrophic forgetting} problem~\cite{Goodfellow14ICLR_Empirical,Kemker18AAAI_Measuring}, i.e., the EI tends to rapidly forget its knowledge of previous domains when trained on the current domain dataset.
This problem severely hinders the EI from gaining the cross-domain sensing capability and, consequently, limits its potential for enabling ubiquitous sensing applications in resource-constrained EDs.

In this paper, we propose \emph{\name}, a resource-efficient framework that enables the EI to continually learn-then-discard domain datasets without catastrophic forgetting, achieving cross-domain continual learning for discriminative sensing tasks in wireless ISAC networks.
The working principle of \name is illustrated in Fig.~\ref{fig_sys_mod}.
Treating EI as a parameterized neural model, we formulate its continual learning over a sequence of domain datasets as a sequence of optimizations of neural parameters.
\name jointly considers the loss on the current domain dataset and a regularization term employed to preserve the knowledge that the EI has learned from previous domain datasets.
After each training, the knowledge is updated, and the used domain dataset is discarded to free up memory occupation, preparing the system to learn subsequent domains.
In particular, in stark contrast with existing studies~\cite{Hu23ICC_Digital, Ji22MobiSys_SiFall} that consider continual learning of EI in ISAC but focus only on the adaptation to the newest domain, \name aims to achieve high sensing accuracy across \emph{all} domains.
As far as the authors know, there are no existing proposals for this important problem in wireless ISAC networks.

Developing \name involves addressing three major challenges.
\emph{Firstly}, unlike video or radar images, the low spatial resolution of EDs and the multipath effects of wireless transmissions make the relationship between CSI and user activities highly intricate, which is further complicated by the nonequispaced sampling time, requiring specialized neural model design.
\emph{Secondly}, both the knowledge and its corresponding regularization term are essential for cross-domain continual learning, yet they are challenging to derive. 
In addition, they must be memory and computational efficient to be applicable given the limited resources of EDs.
\emph{Thirdly}, the training on subsequent domain datasets inevitably leads to deviation of neural parameters from the optimal point for the current domain, degrading its sensing accuracy.
However, such deviations are highly unpredictable, making it difficult to mitigate the accuracy degradation.

To handle the above challenges efficiently, we first design a transformer-based discriminative neural model capable of handling sequences of noisy and nonequispaced CSI samples, which has a compact architecture and relies on no complicated time-frequency transformations.
Then, we propose a regularization term for knowledge retention based on distilled core-sets of few exemplar CSI data, which are significantly smaller than the domain dataset, and improve their efficacy by a novel hybrid clustering-herding exemplar selection method.
Finally, we prevent the neural model from rapid performance loss by endowing it with enhanced robustness towards worst-case parameter deviations.
Our main contributions are:
\begin{itemize}[leftmargin=*]
    \item We propose \name, the first cross-domain continual learning framework for resource-limited EDs in wireless ISAC networks, enabling the EI to learn across a sequence of domain datasets without catastrophic forgetting.
    \item We design an efficient algorithm to handle nonequispaced noisy CSI samples, preserve previous knowledge by replaying distilled core-sets of exemplars, and enhance robustness against unpredictable parameter deviations.
    \item We perform practical experimental evaluations using a real-world wireless ISAC system for HAR, confirming that \name significantly improves the cross-domain capability of EI, achieving 89\% of the cross-domain training accuracy of cumulative training while consuming only 3\% of its memory, mitigating catastrophic forgetting by 79\%.
    \end{itemize}

    The rest of this paper is organized as follows: In Section~\ref{sec_sysmod}, we establish the system model, including the models for CSI and EI in wireless ISAC networks.
    We formulate the \name framework for the cross-domain continual learning of EI in Section~\ref{sec_prob}, and design the algorithm to solve it in Section~\ref{sec_algorithm}.
    Experimental setups and results are presented in Section~\ref{sec_eva}.
    Finally, a conclusion is drawn in Section~\ref{sec_conclu}.

%% file: sections/3_sysmodel.tex
\section{System Model}\label{sec_sysmod}

We consider a system comprising an ED hosting a wireless network for groups of users, where each user connects to the network with its own user device~(UD) carried in close proximity.
The wireless network possesses ISAC capability as the ED uses the uplink CSI to sense each user.
Here, we focus on general discriminative sensing tasks as they are one of the most common tasks in ISAC sensing services.
Without loss of generality, we refer to the sensing target uniformly as \emph{user activity}, regardless of its actual physical meaning.

With its limited computational and memory resources, the ED trains an EI to learn the intricate relationship between CSI variations and user activities. 
However, due to notoriously strong multi-path interference in wireless channels, this relationship, referred to as a \emph{domain}, can be highly specific to surrounding environment and user characteristics.
As a result, the EI tends to be domain-specific and face difficulties in applying to other domains.
\Copy{E-1-2}{\frev{For the considered system, users are designated as the domains since users change while the ED is fixed in an environment, and user characteristics have a determinant impact on the CSI-activity relationship, which will be demonstrated in Sec.~\ref{s3ec_influence}.}}
\Copy{R1-2}{\frev{In other ISAC systems, domains can be represented differently.
For example, if the UD senses its user by downlink CSI, domains can be the EDs that the UD connects to, whose distinct antenna and frequency band determine the CSI-activity relationship.
Furthermore, for the EI of a mono-static ISAC system as in~\cite{He24TWC_Integrated}, domains can be the environmental conditions, which determine the signal echoes associated with user activities.}}

To endow the EI with cross-domain sensing capability, the most effective method is to train it on sufficient training datasets for target domains, referred to as the domain datasets; yet, this is impractical since the ED has limited memory resources and cannot afford storing all domain datasets.
To facilitate addressing the challenges faced by the EI in wireless ISAC networks, below, we establish a general model of the CSI in wireless transmissions, demonstrating its domain specificity.
In addition, we define the EI model and outline how the domain datasets are collected and used.

\subsection{CSI in Wireless Transmissions}\label{s2ec_channel_model}
\subsubsection{Model of CSI}
In view of the symmetric status of each user and its UD, we hereby focus on the CSI between an arbitrary UD and the ED.
Given an uplink data transmitted by a UD as the Tx, the ED as the Rx obtains a CSI sample by using the channel sounding mechanism~\cite{IEEE_standard, 3GPP_TS_138_211}.
Here, a \emph{CSI sample} comprises the channel gains measured by the ED at a certain time for each pair of Tx-Rx antennas over the transmission bandwidth.
Specifically, for a pair of Tx and Rx antennas, within a time duration $T$ and bandwidth $B$, the CSI of frequency $f$ and time $t$ can be modeled as~\cite{goldsmith2005wireless,Chen23ACS_CDWIFI}:
\beq
\label{equ_csi}
h(f,t) = \big(h_{\rstatic}(f) + h_{\rdyna}(f, t) + h_{\rU}(f, t) + \epsilon(f, t)\big) e^{\iu \varphi(f,t)},
\eeq
where $h_{\rU}(f, t)\in\mathbb C$ denotes the channel gain for the dynamic scattering paths relating to the movements of the UD's user, which lays the foundation of sensing the user's activities. 
In contrast, the other terms in~\eqref{equ_csi} constitute interference for the sensing:
$h_{\rstatic}(f)\in\mathbb C$ represents the channel gain for static paths;
$h_{\rdyna}(f, t)\in\mathbb C$ represents the channel gain resulting from user-independent dynamic paths, such as scattering on moving surroundings or on other users;
and $\epsilon(f,t)\in\mathbb C$ denotes the noise that exists inherently in wireless channels.

Besides, in~\eqref{equ_csi}, the phase shift $\varphi(f,t)\in\mathbb R$ represents the random phase error caused by the inevitable unsynchronization between wireless transceivers, resulting from carrier frequency offset, sampling time offset, packet detection delay, etc~\cite{Chen23ACS_CDWIFI}.
Theoretically, $\varphi(f,t)$ can be modeled as a random variable following a uniform distribution within $[0,2\pi)$, $h_{\rstatic}(f)$ can be modeled as a constant complex number, and $\epsilon(f,t)$ can be modeled as a small random variable following complex Gaussian distribution. 
Moreover, based on~\cite{goldsmith2005wireless}, $h_{\rU}(f, t) $ and $h_{\rdyna}(f, t)$ can be modeled approximately by the sum of gains of $S$ scattering paths.
We can focus on the paths experience single scattering since multiple scattering causes significant attenuation.
Therefore, $h_{\rU}(f, t) $ can be expressed as
\begin{align}
    \label{equ_user_gain}
    & h_{\rU}(f,t) =\sum_{s=1}^S   \frac{\lambda \sqrt{G^{(s)}\alpha_{\rU}^{(s)}}\cdot e^{-\iu 2\pi (d_{\rTx, \rU}^{(s)}(t)+d_{\rRx, \rU}^{(s)}(t))/\lambda}}{ (4\pi)^{3/2} d_{\rTx, \rU}^{(s)}(t)\cdot d_{\rRx, \rU}^{(s)}(t) },
\end{align}
where $s$ is the index of the scattering paths, $\lambda = c/f$ is the wavelength of frequency $f$ with $c$ being the light speed, and $\iu$ is the imaginary unit.
Moreover, ${G^{(s)}}$ and $\alpha_{\rU}^{(s)}$ denote the antenna gain and radar cross section for the scattering path, respectively, and $d_{\rTx, \rU}^{(s)}(t)$ and $d_{\rRx, \rU}^{(s)}(t)$ denote the distance from the Tx and Rx antennas to the scattering point of the $s$-th path at time $t$.
Similar to $h_{\rU}(f, t)$, channel gain $h_{\rdyna}(f, t)$ can be modeled by~\eqref{equ_user_gain} with subscript ``$\rU$'' substituted by ``$\rdyna$''.

\subsubsection{Analysis of Domain Specificity} \label{s3ec_influence}
From~\eqref{equ_csi} and~\eqref{equ_user_gain}, it can be observed that sensing the user activity with CSI is susceptible to multiple of interference and noise.
Although the phase error $\varphi(f,t)$ can be handled by conjugate multiplication~\cite{Li17Ubicomp_IndoTrack} (details in Sec.~\ref{sec_alg_1}), and the static gain $h_{\rstatic}(f)$ and the noise $\epsilon(f, t)$ can be handled through band-pass filtering, it is difficult to separate between $h_{\rdyna}(f,t)$ and $h_{\rU}(f,t)$.
This is because the other users' movements are similar to the target user's movements, resulting in similar channel variations as indicated by~\eqref{equ_user_gain}.
Fortunately, in the considered daily scenarios where users keep their UDs in a close distance~(e.g., 20\!~cm), obtaining the CSI variations of the target user is feasible.
Specifically, based on~\cite[Eqn. (4)]{Hu2023Mobicom_Muse}, the following relationship between the CSI variation rates of  $ h_{\rU}^{(s)}$ and $ h_{\rdyna}^{(s)}$ can be derived
\beq
\label{equ_powerRatio_prop}
\frac{|\partial h_{\rU}^{(s)}/\partial t|}{|\partial h_{\rdyna}^{(s)}/\partial t|} \propto
    \Big(\frac{d_{\rTx, \rU}^{(s)}(t)}{d_{\rTx, \rdyna}^{(s)}(t)}\Big)^{-1},
\eeq
where $h_{\rU}^{(s)}$ and $h_{\rdyna}^{(s)}$ denote the gain of the $s$-th path of $h_{\rU}(f,t)$ and $h_{\rdyna}(f,t)$, respectively.
Therefore, since the Tx antenna of the UD is closer to the user while further away from the surroundings and other users, then $d_{\rTx, \rU}^{(s)}(t)\!\ll\! d_{\rTx, \rdyna}^{(s)}(t)$, indicating that the CSI variations caused by the user's movements dominate that caused by other users. 

\Copy{R1-1-1}{Although the close proximity between user and its UD helps separate the CSI variations caused by the user from those induced by other dynamics, it leads to higher domain specificity of the CSI-activity relationship \frev{compared with when the user is far from its UD.}} 
From~\eqref{equ_user_gain}, it is evident that the mapping between CSI and user activity largely depends on the detailed positions and scattering characteristics of the user's scattering points.
\Copy{R1-1-2}{Therefore, for CSI-based sensing in close proximity to the UD, subtle nuances in user behavior have a more pronounced impact \frev{compared with non-proximity sensing.}}  
The following corollary provides a theoretical explanation.

\input{sections/analyses/collary_1.tex}

\Copy{R1-1-3}{Eqn.~\eqref{equ_pos_influence} indicates that when a user is closer to its UD, the CSI variation rate changes more rapidly with the deviation of the scattering point \frev{than when the user is farther away.}} 
Although this implies a higher sensitivity, it also means that the discrepancy among the users' contours and subtle movements will be more influential.
In summary, in scenarios where users are in close proximity to their UDs, the resulting CSI variations can effectively capture the user's activity with minimal interference from other users or environmental dynamics. 
However, this comes at the cost of increased domain specificity of the CSI-activity mapping to individual user nuances.

\subsection{Edge Intelligence in Wireless ISAC Network}\label{s2ec_protocol}

Without loss of generality, an EI can be represented as a neural model, which in essential is a parameterized mapping $\bfg(\cdot ; \bm \theta): \mathcal H \rightarrow \mathcal P$.
Here, $\bfg$ represents the architecture of the neural model, and $\bm\theta\in\mathbb R^{V}$ is its trainable parameters of size $V$. 
In wireless ISAC networks, the input space $\mathcal H$ is the space of time series of CSI samples within duration $T$, each of which is represented by a matrix $\bm H$ referred to as \emph{CSI data}, i.e.,
\beq
\label{equ_csi_data}
\mathcal H \!=\! \{ \bm H\!\in\!\mathbb C^{N\!\times\! (1+L_{\rH})} | ~[\bm H]_{n} \!=\! (t[n], \bm h[n]), n\!=\!1,...,N\!\}, 
\eeq
where $N$ and $n$ denote the number of CSI samples and its index, respectively, 
and $t[n]\in [0,T]$ is the sampling time corresponding to the $n$-th data packet.
We note that the sequence of $t[n]$ is generally nonequispaced according to practical data traffic.
Vector $\bm h[n]\in\mathbb C^{L_{\rH}}$ denotes the $n$-th CSI sample, comprising $L_{\rH}$ elements, and each element represents a channel gain measurement of a certain subcarrier and a Tx-Rx link as defined in~\eqref{equ_csi}.
For the output space $\mathcal P$, it comprises probability vectors for $C$ classes of user activities, i.e.,
\beq
\label{eq_cP}
\mathcal P = \{\bm p\in \mathbb R^C| \sum_{j=1}^C p_j=1,~\bm 0\preceq\bm p\preceq\bm 1\},
\eeq
\Copy{R3-3}{\frev{where the number of classes $C$ is a predetermined integer and known to the EI.}}

Therefore, the sensing of the EI is equivalent to processing $\bm H$ by $\bm p = \bfg(\bm H; \bm \theta)$, where both the architecture $\bfg$ and the parameter vector $\btheta$ play a crucial role.
However, determining them analytically is prohibitively challenging due to the highly intricate relationship between $\bm H$ and $\bm p$, which cannot be addressed in closed-form. 
In contrast, deep learning techniques leverage a general neural network architecture and optimize parameters on training datasets, offering a powerful solution that achieves superior accuracy.

In particular, a training dataset for the EI comprises CSI data and activity labels sampled following the distribution $\varGamma$ of a certain domain, i.e., $\mathcal D=\{ (\bm H_m, \hat{\bm p}_m)\csim \varGamma\}_{m=1}^M$, and thus we refer to $\mathcal D$ as \emph{domain dataset}.
Here, $\hat{\bm p}_m$ represents class label of $\bm H_m$ in the form of a one-hot vector, $M$ denotes the size of domain dataset.
Owing to the significant influence of users' nuances on the CSI as analyzed in Sec.~\ref{s3ec_influence}, the domain is highly dependent on the users whom the CSI data are associated with.
Notably, a neural model is only statistically guaranteed to perform well for the domain of its training dataset, and using it in a different domain, i.e., for a different user, generally leads to poor performance~\cite{goodfellow2016deep}.
Therefore, to develop cross-domain sensing capability and ensure high sensing performance across all users, the EI must be trained on various domain-specific datasets.

However, due to the ED's limited memory resources, storing all domain datasets is impractical. 
For instance, with a sampling rate of 100 packets per second, an 80 MHz bandwidth, and $2 \times 2$ Tx-Rx pairs, a 20-activity dataset for 20 users will exceed 100 GB if each activity involves just 5 minutes of CSI data. 
This imposes a significant storage burden on the ED and results in substantial overhead if the data is uploaded to a cloud server. 
Moreover, as these datasets are collected over extended periods, the need for periodic retraining to incorporate new domains adds a considerable computational load.

To alleviate such heavy memory burdens, a potential approach is for the EI to sequentially learn the domain datasets collected over different time periods. 
For clarity, we assume that in the $k$-th~($k=1,\dots,K$) period, a domain dataset $\mathcal{D}_k$ is collected. 
The EI then updates the neural model's parameters $\btheta$ using $\mathcal{D}_k$ before discarding the dataset to free up memory. 
We note that, without loss of generality, a domain dataset may contain labeled CSI data from multiple users, while different domain datasets correspond to different users.
Although this approach alleviates storage pressure, it introduces the challenge of catastrophic forgetting during sequential learning. 
In the following section, we formally formulate this issue and propose the \name framework to address it.

%% file: sections/analyses/collary_1.tex
\begin{corollary}
\label{corollary_1}
Denote by $h$ the channel gain for a single scattering path as defined in~\eqref{equ_user_gain}.
In addition, denote the distance between the UD's Tx antenna to the scattering point by $d$.
Then, the influence of $d$ on the relative variation rate of $h$ is inversely proportional to $d$, i.e.,
\beq
\label{equ_pos_influence}
\frac{\partial }{\partial d}\Big(\Big| \frac{\partial h}{\partial t}\Big|\Big) /  \Big| \frac{\partial h}{\partial t}\Big|\approx -\frac{1}{d}.
\eeq
\end{corollary}
\begin{IEEEproof}
Eqn.~\eqref{equ_pos_influence} can be derived by first taking partial derivative of $d$ on the root of~\cite[Eqn. (4)]{Hu2023Mobicom_Muse} and then calculating the relative ratio.
\end{IEEEproof}

%% file: sections/4_problem.tex
\section{\name: Cross-Domain Continual Learning for Edge Intelligence of ISAC}\label{sec_prob}

We propose the \name framework to handle the catastrophic forgetting challenge by optimizing the cross-domain sensing accuracy of the EI during it continually learns a sequence of domain datasets $\cD_1,...,\cD_K$.
At each period $k$, \name aims to minimize the empirical classification loss of the EI across the current and previous domain datasets $\cD_1,...\cD_{k}$.
The optimization problem can be expressed as the following sequential training problem for parameter $\btheta$:
\begin{align}
\text{\textbf{(P0)}}_k\quad\min_{\btheta}~&\cC(\btheta;\cD_{1:k}), \nonumber\\
\text{s.t.}~&\cC(\btheta;\cD_{1:k}) =\!\sum_{(\bm H, \hat{\bm p} )\in\cD_{1:k}} \!{\ell (\bgun(\bm H;\btheta), \hat{\bm p})}, \nonumber
\end{align}
where $\mathcal D_{1:k}=\cup_{i=1}^k \mathcal D_i$ denotes the union set of $\cD_1,...,\cD_k$,
and $\ell(\bm p, \hat{\bm p})$ represents the cross-entropy~(CE) loss given the sensing result of the EI and the ground truth one-hot label.
More specifically, the CE loss can be expressed~as
\beq
\ell(\bm p, \hat{\bm p}) = - \sum_{j=1}^C \hat{p}_j \log (p_j).
\eeq

However, the direct formulation of~(P0)$_k$ is intractable because at the $k$-th period, the previous domain datasets $\cD_1, ...\cD_{k-1}$ have already been discarded.
To convert (P0)$_k$ into a tractable form, we reformulate it as an equivalent problem: Solving the parameter $\btheta$ with the maximum \emph{a posteriori} probability given domain datasets $\cD_1,...,\cD_k$, i.e.,
\beq
\label{equ_map}
\btheta^* = \arg\max_{\bm\theta}~\log(\Pr(\bm\theta|\cD_{1:k})).
\eeq
Based on Bayes' rule, the objective in~\eqref{equ_map} is derived as:
\begin{align}
\label{equ_bayesian_equ}
\log(&\Pr(\bm\theta|\cD_{1:k})) =  \\
& \log(\Pr(\cD_{k}|\bm\theta)) +  \log(\Pr(\bm\theta|\cD_{1:k-1})) - \log(\Pr(\mathcal D_k)), \nonumber
\end{align}
where $\log(\Pr(\cD_{k}|\bm\theta))$ is the log-likelihood function of $\mathcal D_k$ given parameter $\bm \theta$, which is negatively proportional to the CE loss over $\mathcal D_k$, i.e., $\log(\Pr(\cD_{k}|\bm\theta))\propto-\cC(\btheta;\cD_k)$.
Since $\Pr(\cD_k)$ is independent of $\btheta$, \eqref{equ_bayesian_equ} reveals that (P0)$_k$ can be handled without directly accessing $\cD_{1:k-1}$ but by leveraging \emph{a posteriori} probability distribution $\Pr(\bm\theta|\cD_{1:k-1})$. 

As the exact \emph{a posterior} probability distribution can hardly be obtained in practice, it is necessary to adopt an approximation of $- \log (\Pr(\bm\theta|\cD_{1:k-1}))$.
Without loss of generality, we denote the approximation by function $\mathcal R(\bm\theta; \cK_{k})$, with $\cK_{k}$ representing a core-set of knowledge from the domains preceding the $k$-th domain.
Consequently, (P0)$_k$ can be converted to
\begin{align}
    \text{\textbf{(P1)}}_k \quad \min_{\bm \theta}~ &\cL_k(\btheta) = \mathcal R(\bm \theta; \cK_{k}) + \mathcal C(\btheta; \cD_k). \nonumber 
\end{align}

In the converted problem (P1)$_k$, $\cR(\btheta;\cK_{k})$ serves as \emph{regularization}, incorporating knowledge of previous domains into the training of $\btheta$ over $\cD_k$, thereby enabling continual learning across the $k$ domains and preventing catastrophic forgetting.
Therefore, we formulate the cross-domain continual learning of the EI as a sequence of optimization problems, i.e., (P1)$_k$ ($k=1,\dots,K$). 
It can be seen that the key to effectively solving (P1)$_k$ is to determine the appropriate regularization $\cR(\btheta;\cK_{k})$ as well as the core-set of knowledge $\cK_{k}$.
Furthermore, owing to the limited resources of the ED, solving (P1)$_k$ is subject to the following two implicit constraints:
\begin{itemize}[leftmargin=*]
\item \textbf{Memory constraint}: In line with the goal of alleviating the memory burden of ED, the memory consumption of $\cK_{k}$ should be significantly lower than that of $\cD_{1:k-1}$.
\item \textbf{Computation constraint}: Given the limited computational resources of the ED, the algorithmic complexity for deriving $\cK_{k}$ must be kept low.
\end{itemize}

Solving (P1)$_k$ under the memory and computation constraints faces three critical challenges:
\emph{First}, for the neural parameters to map CSI data into accurate user activities, their architecture needs to be carefully designed, which is challenging due to the difficulty in handling nonequispaced CSI samples affected by noise and interference.
\emph{Second}, due to the complexity of the high-dimensional \emph{a posteriori} probability distribution $\Pr(\bm \theta| \cD_{1:k-1})$, determining the regularization function that can effectively approximate it and prevent catastrophic forgetting is a major challenge, which is exacerbated by the resource constraints of the ED. 
\emph{Third}, since subsequent training following (P1)$_k$ will cause deviations in the optimized neural parameters, it is challenging to solve (P1)$_k$ with robustness against such unpredictable changes.

%% file: sections/5_algorithm.tex
\section{Algorithm Design}\label{sec_algorithm}

\begin{figure}[t]
    \centering
    \includegraphics[width=.95\linewidth]{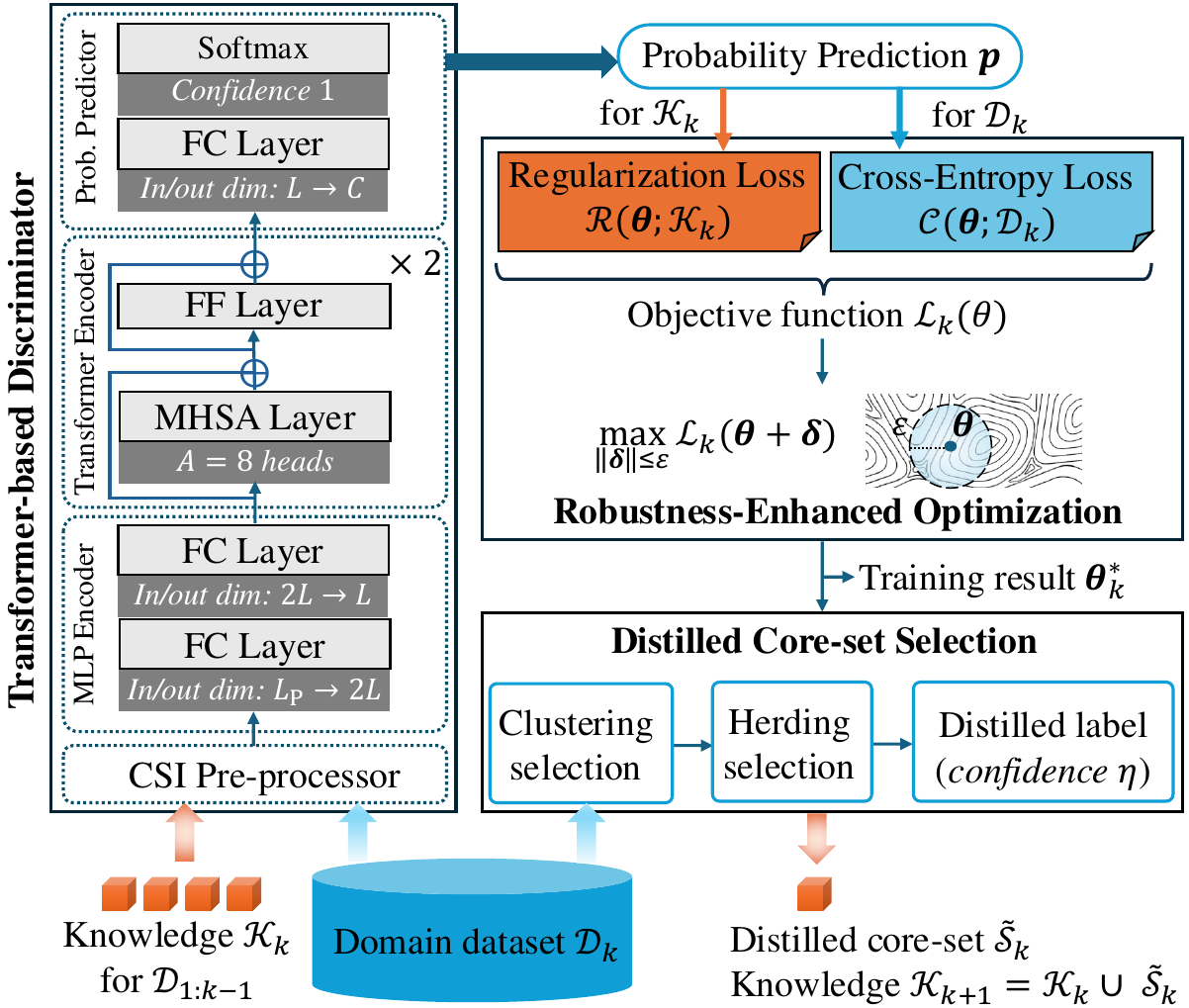}
    \vspace{.5em}
        \caption{\Copy{cap_new_fig2}{\frev{Diagram of the proposed algorithm, including the transformer-based discriminator, the robustness-enhanced optimization, and the distilled core-set selection.}}}
        \label{fig_alg_illu}
\vspace{-.5em}
\end{figure}

We propose an efficient algorithm for \name to handle the cross-domain continual learning problem (P1)$_k$ of the EI.
As illustrated in Fig.~\ref{fig_alg_illu}, the algorithm comprises three parts:
We first design the neural model as a transformer-based discriminator with compact architecture, handling the first challenge of (P1)$_k$.
Then, to handle the second challenge, we derive a resource-efficient regularization function and corresponding knowledge to prevent catastrophic forgetting in training by replaying distilled core-sets.
Furthermore, for the third challenge, we leverage robustness-enhanced optimization to reduce performance degradation caused by worst-case deviations. 
\Copy{E-1-3}{\frev{We note that although we designate domains as users due to their dominant impacts on the CSI-activity relationship, our proposed algorithm is independent of domain designation and thus effective for arbitrary designation of domains in various wireless ISAC systems.}}

 \subsection{Transformer-Based Discriminator for CSI Sequences}\label{sec_alg_1}
We design the architecture of the neural model to comprise two parts: a CSI pre-processor and transformer-based sequence classifier. 
The transformer module is adopted to handle CSI data because CSI data are composed of sequences of CSI samples, and transformers are universal approximators for arbitrary sequence functions~\cite{Yun2020ICLR_Are}.
Moreover, based on~\cite{Hu24JSAC_Cross}, transformers can handle nonequispaced time series efficiently.
In the following, we describe the key components of the CSI pre-processor and the sequence classifier, respectively.

\subsubsection{CSI Pre-processor}
The pre-processor serves two purposes. 
Firstly, it expands each scalar sampling time into a temporal feature vector to help capture temporal relationship among CSI samples.
Secondly, it processes each complex CSI sample into a real-valued vector, suppressing random phase errors in~\eqref{equ_csi} by conjugate multiplication and normalization.

In particular, for the $n$-th row of $\bm H$ ($n=1,\dots,N$), i.e., $[\bm H]_n=(t[n], \bm h[n])$ defined in~\eqref{equ_csi_data}, we expand the scalar time $t[n]$ into a vector $\bm \tau[n] \in \mathbb C^{L_{\rT}}$ by leveraging a series of periodic trigonometric functions.
This enhances the representation of temporal information~\cite{Wen23IJCAI_Transformer}, allowing the neural model to better capture detailed temporal relationships between CSI samples, particularly the periodicity of phase variations at different speeds.
More specifically, the $j$-th ($j=1,\cdots, L_{\rT}$) element of temporal feature vector $\bm\tau[n]$ can be expressed as
\beq
\label{equ_time_embed}
\tau_j[n] = \begin{cases}
    \sin(t[n]/ T^{j/L_{\rT}}), &\text{if $j$ is even,}\\[-.0em]
    \cos(t[n]/ T^{(j-1)/L_{\rT}}),&\text{otherwise,}
\end{cases}
\eeq
where $T$ denotes the maximum time duration that the CSI data spans, i.e., $t[N]\leq T$.

Subsequently, we handle the $n$-th CSI sample, i.e., $\bm h[n]$, which comprises the channel gains between each pair of Tx and Rx antennas for all subcarriers.
To handle the random phase error in~\eqref{equ_csi}, we adopt the conjugate multiplication method in~\cite{Li17Ubicomp_IndoTrack}, which multiplies the conjugate channel gain of the first Rx antenna with the channel gains of the other antennas.
This approach effectively suppresses random phase errors, as the RF chains of different Rx antennas share a common oscillator, causing their random phase errors on the CSI to be similar.
Moreover, we normalize the conjugate multiplication result by detrending and then dividing it by the maximum amplitude.
Therefore, for an arbitrary Tx antenna and subcarrier with their indexes omitted for clarity, the $n$-th pre-processed CSI sample comprises
\beq
\label{equ_x_data}
x_r[n] = \norm(h_{r+1}[n]\cdot \overline{h_1}[n]),\quad r=1,\dots,N_{\rRx}-1,
\eeq
where $r$ is the index of the Rx antenna, 
$N_{\rRx}$ denotes the total number of Rx antennas,
$h_r[n]$ and $h_1[n]$ denote the elements of $\bm h[n]$ corresponding to the CSI at the $r$-th and the first Rx antennas, respectively, 
and $\norm(\cdot)$ represents the aforementioned normalization.
The efficacy of the CSI pre-processing in~\eqref{equ_x_data} is validated by Proposition~\ref{prop_2_x_data} below.

\input{sections/analyses/prop_2.tex}

Proposition~\ref{prop_2_x_data} indicates that the pre-processed CSI holds the information of the channel changes caused by the user's movements in proximity to the UD.
We note that in~\eqref{equ_efficacy_x}, \frev{$\beta_r \overline{\Delta h_{\rU,1}}[n]$} is environment dependent, which may hinder the clear correlation between $\Delta x_r[n]$ and $\Delta h_{\rU,r}[n]$.
Nevertheless, its impact can be less significant when the $r$-th Rx antenna is closer to the first Rx antenna.
In that case, since $|h_{\rstatic,r}|\approx |h_{\rstatic,1}|$ and $|\Delta h_{\rU,r}[n]|\approx |\Delta h_{\rU,1}[n]|$, based on~\eqref{equ_efficacy_x} and the Euler equation, it can be derived that 
\begin{align*}
    \Delta x_r[n] \! \propto\! e^{\iu \varphi_r[n]} \cdot \big|\Delta h_{\rU,r}[n]\big|\cdot\cos\!\big(\angle(\Delta h_{\rU,r}[n])-\varphi_r[n]\big),
\end{align*}
where $\varphi_r[n] = \angle(\beta_r\cdot \Delta h_{\rU, r}[n] / \Delta h_{\rU, 1}[n])/2$.
As the phase of channel gain generally changes much faster than its amplitude, it can be observed that $|\Delta h_{\rU,r}[n]|$ can be inferred by the envelope of $|\Delta x_r[n]|$, and $\angle(\Delta h_{\rU,r}[n])$ can be inferred by subtracting cosine phase of $|\Delta x_r[n]|$ by $\angle(\Delta x_r[n])$.
This demonstrates the feasibility of obtaining channel variations caused by the user's movements from the CSI samples after the pre-processing in~\eqref{equ_x_data}.

In more complex scenarios where the Rx antennas are separated more widely, the relationship between $\Delta h_{\rU,r}[n]$ and $\Delta x_{r}[n]$ can be more intricate, and thus an efficient neural model is needed by the EI.
As neural models are more efficient in handling real values instead of complex values, we convert each complex-valued input element $x\in\mathbb C$ into its equivalent real-valued vector $(|x|, \cos\angle x, \sin\angle x)$ as in~\cite{Hu24JSAC_Cross}.
In effect, this complex-to-real conversion preserves the cyclicity of $\angle x$ and avoids the discontinuity between $0$ and $2\pi$.

Concatenating each temporal feature vector in~\eqref{equ_time_embed} and the corresponding pre-processed real-valued vector in~\eqref{equ_x_data}, the resulting sequence can be arranged as a matrix:
\beq
[\bm X]_n = \big(\bm\tau[n], |\bm x[n]|, \cos(\angle\bm x[n]), \sin(\angle\bm x[n])\big),
\eeq
where $n=1,\dots,N$, $\bm X\in\mathbb R^{N\times L_{\rP}}$ with $L_{\rP}$ denoting the vector length after pre-processing, and $\bm x[n]$ comprises the pre-processed CSI for all subcarriers, and Tx-Rx antenna pairs.

\subsubsection{Sequence Classifier}\label{s3ec_seq_class}
Considering the limited computational resource of the ED, we avoid complicated architectures with massive parameters and design the following compact architecture, which comprises three modules: a multi-layer perceptron~(MLP) encoder, a transformer encoder, and a fully-connected~(FC) classifier.
The details of the architecture are illustrated in the left part of Fig.~\ref{fig_alg_illu}.
Ass MLPs and transformers are proven universal approximators for arbitrary functions~\cite{Chen1995Universal,Yun2020ICLR_Are}, the proposed architecture is expected to achieve high performance for general CSI-activity mapping.

\textbf{MLP Encoder}: To extract activity-related features from each pre-processed sample, i.e., $[\bm X]_n$, the MLP encoder maps each sample into a feature vector, leveraging two cascaded FC layers.
\Copy{R3-2}{\frev{The first FC layer takes each $[\bm X]_n$ ($n=1,...,N$) as input and outputs a feature vector of $2L$ dimensions. The second FC layer takes the output feature vectors from the first layer, mapping each of them to an output of $L$ dimensions.}}
In particular, given an input feature vector $\bm x_{\rin}\in \mathbb R^{L_{\rin}}$, each FC layer can be expressed as a function below:
\beq\label{equ_fc_expression}
\mathtt{FC}(\bm x_{\rin}) = \mathtt{Norm}\big(\mathtt{ReLU}(\bm W\bm x_{\rin}+\bm b)\big),
\eeq
where $\bm W\in\mathbb R^{L_{\rout}\times L_{\rin}}$ and $\bm b \!\in\! \mathbb R^{L_{\rout}}$ denote the trainable weight and bias parameters, respectively,
$\mathtt{ReLU}(\cdot)$ is the ReLU activation function,
and $\mathtt{Norm}(\cdot)$ represents normalizing to zero mean and unit variance.
\Copy{R3-2c-1}{\frev{Collecting the $N$ output vectors of $L$ dimensions from the second FC layer, the MLP encoder arranges them as a new sequence denoted by $\bm Z\in\mathbb R^{N\times L}$, which is then fed into the transformer encoder.}}

\textbf{Transformer Encoder}
The transformer encoder comprises cascaded multi-head self-attention~(MHSA) layers and feed forward~(FF) layers, where each layer is equipped with a residual connection~(RC)~\cite{He2016_CVPR} to mitigate the vanishing gradient problem in training a deep neural model.
The core of the transformer encoder is using MHSA layers to enable the information in individual feature vectors of a sequence to complement each other according to their correlation.
\Copy{R3-2c-2}{\frev{The first MHSA layer is fed with the sequence $\bm Z\in\mathbb R^{N\times L}$ obtained from the MLP encoder.}}
It splits each input vector to apply multiple self-attention layers and then concatenates the outputs, which can be expressed as
\begin{align}
& \mathtt{MHSA}(\bm Z) = \tAttn_{1}(\bm Z_1)\oplus \dots \oplus \tAttn_{A}(\bm Z_{A}), \nonumber \\
& \tAttn_{a}(\bm Z_a)  = \mathtt{softmax}\Big( \frac{\bm Z_a\bm W_{\rQ,a} (\bm Z_a \bm W_{\rK,a})^\tp}{\sqrt{L/A}}
    \Big) \cdot (\bm Z_a \bm W_{\rV,a}),\nonumber 
\end{align}
where $\bm Z_a\in\mathbb R^{N\times L/A}$ is the $a$-th ($a=1,\dots,A$) sub-matrix of $\bm Z$ with $\bm Z = \bm Z_1\oplus\dots\oplus\bm Z_A$,
and $\bm W_{\rQ,a},\bm W_{\rK,a},\bm W_{\rV,a}\!\in\!\mathbb R^{L/A \times L/A }$, are trainable parameters referred to as the query, key, and value matrices, respectively.
\Copy{R3-2c-3}{\frev{The output of the MHSA layer is added with its input through the RC, which can be expressed as 
\beq
\tilde{\bm Z} = \bm Z + \mathtt{MHSA}(\bm Z). \nonumber
\eeq
Then, $\tilde{\bm Z}$ is sent to the FF layer, which comprises two FC layers with equal dimensions of input and output.
The FF layer handles each row vector $\tilde{\bm z}\in\mathbb R^{L}$ in $\tilde{\bm Z}$ by
\beq
\mathtt{FF}(\tilde{\bm z}) = \mathtt{FC}_2(\mathtt{FC}_1(\tilde{\bm z})),
\eeq
which refines feature representations by feature-level interactions.
The output of the FF layer is added with its input through the RC, which is then fed into the next MHSA layer.}}

\textbf{Probability Predictor}:
As the feature vectors are mutually complemented in the transformer encoder, we can use the first feature vector in the sequence as the input for the final classification, which reduces the computational cost.
This input feature vector is handled by an FC layer to reduce its dimension to the number of classes $C$, the resulting logits are denoted by vector $\bm y\in\mathbb R^C$.
Finally, a softmax function is employed to convert the logit vector into probability predictions for the $C$ classes, i.e.,
\beq
\bm p = \mathtt{softmax}(\bm y) =  \frac{\exp(\bm y)}{\sum_{i=1}^C \exp(y_i)}.
\eeq

\subsection{Knowledge Retention by Replaying Distilled Core-Set}\label{s2ec_know_retent}

We then design the regularization function $\mathcal R(\bm\theta; \cK_{k})$ in (P1)$_k$, which should approximate $- \log (\Pr(\bm\theta|\cD_{1:k-1}))$.
In the context of solving sequential training (P1)$_k$, a common approach to approximate $\Pr(\bm\theta|\cD_{1:k-1})$ is using a multivariate Gaussian distribution, with the initial point of (P1)$_k$, or equivalently, the result of (P1)$_{k-1}$, as the mean and variance estimated statistically.
In this case, the regularization function can be expressed as
\beq
\label{equ_R_G}
\cR_{\rP}(\btheta; \cK_{k}) = \frac{1}{2}\sum_{i=1}^V v_{k,i} (\theta_i - \theta_{k-1,i}^*)^2, 
\eeq
where core-set knowledge $\cK_{k}\!=\!\{\btheta^*_{k-1}, \bm v_{k}\}$, $\btheta^*_{k-1}\!=\!(\theta^*_{k-1,1},\dots,\theta^*_{k-1,V})$ denotes the result of solving (P1)$_{k-1}$, and $\bm v_{k} \!=\!(v_{k,1},\dots,v_{k,V})$ denotes the variance vector.
Due to its intuitive interpretation of penalizing deviations in important parameters for knowledge retention, such approximation has been widely used in mainstream continual learning studies~\cite{Kirkpatrick17PNAS_Overcome, Aljundi18ECCV_Memory,Zhao24ICML_Statistical}, which are referred to as \emph{parameter regularization}~(PR) approaches. 

However, despite their merit in intuitive interpretation and constant memory consumption, the performance of PR approaches is usually unsatisfactory~\cite{Van22NatMI_Three, Delange22PAMI_Continual, Akrout23ICC_Continual} since approximating the \emph{a posteriori} distribution as a Gaussian can be overly simplistic.
As practical $\Pr(\bm\theta|\cD_{1:k-1})$ is highly complex, a simple parametric distribution such as the Gaussian can only approximate it within a rather limited local region.
Yet, during the sequential training, the neural parameters deviate from their initial points to learn new domain datasets.
As a result, the penalty inevitably becomes increasingly less accurate, causing PR to fail in preserving prior knowledge.

For the above reasons, we design the regularization function based on an alternative type of approaches named \emph{experience replay}~(ER)~\cite{Chaudhry19ICML_Continual}, where the approximation of the \emph{a posteriori} distribution is less constrained to a local region.
The principle of ER is to select a small set of exemplars $\cS_{1:k-1}$ from previous domain datasets $\cD_{1:k-1}$ as the core-set of knowledge and replay them during training on $\cD_k$.
Essentially, it is equivalent to applying the following approximation:
\beq
\label{equ_approx_2}
\log\!\big(\Pr(\btheta|\cD_{1:k-1})\big)\approx \log\!\big(\Pr(\btheta|\cS_{1:k-1})\big).
\eeq
When $\cS_{1:k-1}$ is randomly sampled from $\cD_{1:k-1}$ and $|\cS_{1:k-1}|$ is sufficiently large, the approximation in~\eqref{equ_approx_2} is expected to holds for all $\btheta$, rather than restricted to a local region as in PR approaches. 
Here, $|\cdot|$ denotes the cardinality of the given set.
In this regard, by using a set of randomly selected data samples from previous domain datasets as the core-set of knowledge and involving their CE loss in (P1)$_k$ as the regularization term, it seems (P1)$_k$ can be readily handled.

Nevertheless, to satisfy the memory constraint of the ED, the number of selected exemplars needs to be sufficiently small. 
In this case, random selection often results in a highly biased and non-representative core-set, compromising the efficacy of knowledge retention.
Several studies propose using generative models~\cite{Shin17NIPS_Continual} or data condensation techniques~\cite{Liu20CVPR_Mnemonics} to generate core-sets that possess full knowledge of previous datasets, but these methods are unsuitable for the ED due to their high computational complexity.
In contrast, herding~\cite{Rebuffi17CVPR_iCaRL} and clustering-based~\cite{Nguyen18ICLR_VCL} exemplar selection are more computationally efficient while achieving comparable performance~\cite{Van22NatMI_Three}; yet, each of them have its own strengths and limitations.

To enhance the efficiency of selecting representative exemplars, we propose a novel hybrid method to combine the strengths of herding and clustering-based selection while addressing their limitations.
\Copy{R1-5}{\frev{Preliminarily, the clustering-based method selects exemplars whose feature vectors are closest to the cluster centers revealed by the K-means algorithm. The herding-based method, on the other hand, selects exemplars whose mean feature vector is closest to the overall mean of the feature vectors in the domain dataset. The detailed expressions of the two methods will be provided in~\eqref{equ_kmeans} and~\eqref{equ_herding}, respectively.}}

Without loss of generality, we denote the neural parameter after training in the current period by $\btheta^*$, omitting the subscript for period index.
Using the two encoder parts of $\bgun(\cdot;\btheta^*)$, we project the current domain dataset $\cD$ onto its latent feature space in $\mathbb R^L$, yielding $\cF$.
Then, $\cF$ is divided according to the classes, with the subset of class $c$ being denoted by $\cF_{c}$.
For each $\cF_c$, the budget for exemplars is denoted by $E$, with $E\ll |\cF_{c}|$, which is allocated between clustering and herding-based selection according to the ratio $\beta\in[0,1]$, referred to as the \emph{clustering-herding ratio}.

The first $\beta E$ exemplars are selected by using the K-means clustering algorithm, representing $\beta E$ centers of the feature vectors in $\cF_{c}$.
In particular, it is equivalent to finding $\beta E$ \emph{centroid feature vectors} in $\cF_c$ that minimizes their collective distance to the rest of the feature vectors, i.e.,
\beq
\label{equ_kmeans}
\bfv_{c,1}^*,\dots,\bfv_{c, \beta\!E}^* = \!\argmin_{\bfv_j\in\cF_c|_{j=1,\dots,\beta\!E}} \sum_{\bfv\in\cF_c} \min_{j=1,\dots,\beta\!E}(\|\bfv - \bfv_j\|_2^2).
\eeq

Although the above clustering problem is a combinatorial optimization problem and is NP-hard, it can be solved approximately by a greedy algorithm named Lloyd's algorithm~\cite{Kanungo02PAMI_Efficient}.  
The Lloyd's algorithm starts from randomly picking initial cluster centroids and then iterates between clustering feature vectors to their nearest centroids and updating centroids as the means of the clusters. 
In addition, we employ the k-means++ method~\cite{Arthur06kpp} to enhance the random selection of initial centroids, which essentially increases the mutual distance among the selected centroids, leading to better convergence and clustering results.

Nevertheless, solely using the clustering method to select exemplars may result in lack of representation for the center of the entire feature vector set, i.e., $\bar{\bfv_c} = \sum_{\bfv\in\mathcal F_c} \bfv / |\cF_c|$.
Based on~\cite{Gretton12JMLR_Kernel}, minimizing the distance between the mean of feature vectors of the selected exemplars and $\bar{\bfv}$ is important for matching the distributions between the selected exemplars and the original dataset.
To achieve this goal, we employ the herding method~\cite{Rebuffi17CVPR_iCaRL} to select the rest of $(1-\beta)E$ exemplars so that the overall mean of the selected feature vectors has a minimal distance from $\bar{\bfv}_c$.
It is equivalent to solving the following optimization problem:
\begin{align}
\label{equ_herding}
\bfv_{c, \beta\! E+\!1}^*,\dots,\bfv_{c,E}^*  = 
&\argmin_{\bfv_{j}\in\cF_c|_{j=\beta\! E+\!1,\dots,E}} \big\| \sum_{j'=1}^{E}\! \bfv_{j'}- \bar{\bfv}\big\|_2^2 \\
&\qquad~ \text{s.t.~$\bfv_{j'} = \bfv_{c,j'}^*$, $\forall j'=1,\cdots, \beta\!E$.} \nonumber
\end{align}
Again, although~\eqref{equ_herding} is a challenging combinational optimization problem, it can be solved approximately using a greedy algorithm.
Specifically, we employ the nearest-class-mean algorithm in~\cite{Rebuffi17CVPR_iCaRL}, which iteratively finds each feature vector to minimize the current distance.

Consequently, the CSI data corresponding to the $E$ feature vectors obtained by~\eqref{equ_kmeans} and~\eqref{equ_herding} are selected as the exemplars, forming the core-set of class $c$, i.e.,
\beq
\label{equ_ori_coreset}
\cS_c = \{ (\bm H,\hat{\bm p})\in\cD_c | \hat{\bgun}(\bm H;\btheta^*) = \bfv^*_j,~\exists j=1,\dots E\},
\eeq
where $\hat{\bgun}(\cdot)$ denotes the encoder parts of the neural model, i.e., the neural model without the probability predictor.

We then proceed to deriving the regularization function.
Since the empirical CE loss approximates the negative logarithmic \emph{a posterior} distribution, it is intuitive to adopt the CE loss over the core-set as the regularization function.
Nevertheless, due to the small size of the core-set, the precision of is approximation is low.
To enhance the precision, instead of using the one-hot class label $\hat{\bm p}$ of each exemplar, we adopt the probability vector predicted by the trained neural model. 
\Copy{R1-6}{\frev{Furthermore, to mitigate the over-confidence issue~\cite{Guo17ICML_Calibration} of the neural model during training on the small number of exemplars, we divide the logits output associated with the exemplars in $\bgun(\cdot)$  by a confidence downscaling factor $\eta>1$ before the softmax.}}
This design essentially coincides with the \emph{distillation loss}~\cite{Gou21IJCV_Knowledge}, which is proven to enhance the knowledge transfer between models.
In this regard, we refer to the selected core-set with low-confidence probability labels as \emph{distilled core-set}.
For the $k$-th period, the distilled core-set can be expressed as:
\begin{align}
 \tilde{\cS}_{k} =  \bigcup_{c=1}^{C} \!\tilde{\cS}_{k,c}, ~
 \tilde{\cS}_{k,c} = \{ (\bm H, \tilde{\bm p}) | \tilde{\bm p} = \bgun_{\eta}(\bm H; \btheta_{k}^*), \bm H \in \cS_{k,c}\},  \nonumber
\end{align}
where $\cS_{k,c}$ denotes the core-set in~\eqref{equ_ori_coreset} for the $k$-th period,
${\bgun}_{\eta}(\bm H; \btheta)$ denotes the neural model with its confidence being divided by $\eta$,
and $\tilde{\bm p}$ is referred to as a \emph{distilled label}.

Consequently, we propose to achieve the knowledge retention by replaying the distilled core-set during the training, and thus the regularization function can be derived as
\begin{align}
\label{equ_regu_func}
\cR(\btheta; \cK_{k}) &= \sum_{(\bm H, \tilde{\bm p})\in \cK_k} {\ell({\bgun}_{\eta}(\bm H; \btheta), \tilde{\bm p})},
\end{align}
where knowledge core-set $\cK_{k} = \cup_{k'=1}^{k-1}\tilde{\cS}_{k'}$.

\subsection{Robustness-Enhanced Parameter Optimization}\label{s2ec_robustness}
We address the performance degradation due to the unpredictable neural parameter deviation caused by training in subsequent domains.
Intuitively, given the optimized neural parameters at the $k$-th period denoted by $\btheta^*_k$, the training in the subsequent domains will inevitably lead to a deviation $\bm\delta\in\mathbb R^V$ from $\btheta^*_k$, and thus we need to minimize $\cL_k(\btheta^*_k+\bdelta)$ and $\cL_k(\btheta^*_k)$.
Nevertheless, owing to the causality, the deviation $\bm \delta$ is unknown in the training for the $k$-th domain.
To handle this problem, we can solve the $\btheta^*_k$ while minimizing the worst-case loss for all possible deviations from it, which we refer to as the \emph{robustness-enhanced parameter optimization}:
\beq
\label{equ_sam}
\btheta^*_{k} = \arg\min_{\btheta} ~ \max_{\|\bdelta\|_2\leq \epsilon} \cL_{k}(\btheta + \bdelta),
\eeq
where $\epsilon$ denotes the considered \emph{deviation radius} and is generally small ($10^{-2}\!\sim\! 10^{-1}$) to ensure the efficiency of optimizing $\btheta$.
The optimization objective in~\eqref{equ_sam} can be viewed as finding an optimal $\btheta$ where the sharpness of its local loss landscape is minimized~\cite{Foret20ICLR_Sharpness}, which is closely related to the principle of meta-learning~\cite{Finn17ICML_Model}.
By substituting the objective of (P1)$_k$ with that of~\eqref{equ_sam}, the robustness of the resulting $\btheta^*_k$ can be significantly enhanced, enduring deviations with minimal potential increase in loss during subsequent training.

To handle~\eqref{equ_sam} efficiently, we apply first-order Taylor expansion on $\cL_k(\btheta+\bdelta)$ around $\btheta$.
Then, the deviation that maximizes the loss around $\btheta$ can be approximated by
\begin{align}
    \label{equ_sam_delta}
    \bdelta^* & \approx \argmax_{\|\bdelta\|_2\leq \epsilon}~\cL_k(\btheta) + \bdelta^\tp \nabla_{\btheta} \cL_k(\btheta)\\
    \label{equ_sam_delta2}
    & = \epsilon \cdot \nabla_{\btheta} \cL_k(\btheta) / \|\nabla_{\btheta} \cL_k(\btheta)\|_2,
\end{align}
which is an insightful result since the direction of gradient $\nabla_{\btheta} \cL_k(\btheta)$ indicates the direction of the steepest increase of $\cL_k(\btheta)$.
Leveraging~\eqref{equ_sam_delta}, problem~\eqref{equ_sam} can be converted to
\beq
\label{equ_converted_prob}
\btheta^*_{k} = \arg\min_{\btheta} ~\cL_{k}\big(\btheta + \epsilon \cdot  \frac{\nabla_{\btheta}\cL_k(\btheta)}{\|\nabla_{\btheta} \cL_k(\btheta)\|_2}\big).
\eeq

Without loss of generality, we assume that the optimizer employed to solve the converted problem~\eqref{equ_converted_prob} is gradient-based.
In this case, the gradient of the objective function in~\eqref{equ_converted_prob} can be approximately calculated by
\begin{align}
    \label{equ_sam_grad}
    \nabla_{\btheta} \cL_k(\btheta+\bdelta^*) &\approx  \nabla_{\btheta} \cL_k(\btheta)|_{\btheta = \btheta+\bdelta^*}\\ 
    & + \epsilon \cdot \left( \nabla_{\btheta}\frac{ \nabla_{\btheta} \cL_k(\btheta) }{\|\nabla_{\btheta} \cL_k(\btheta)\|_2}\right)^\tp\nabla_{\btheta} \cL_k(\btheta)|_{\btheta = \btheta+\bdelta^*}. \nonumber 
\end{align}

On the right hand side of~\eqref{equ_sam_grad}, the first term is the gradient of loss function at $\btheta+\bdelta^*$ instead of $\btheta$, and the second term represents the compensation for the change of gradient direction from $\btheta$ to $\btheta+\bdelta^*$, which is generally small due to small $\epsilon$. 
However, calculating the second term requires the estimation of the Hessian matrix of $\cL_k(\btheta)$, which can result in substantial computational burden for the ED.
Besides, as shown in~\cite{Foret20ICLR_Sharpness}, involving high-order derivatives of $\cL_k(\btheta)$ may degrade the performance for estimating $\nabla_{\btheta} \cL_k(\btheta+\bdelta^*)$ due to more numerical errors.
Therefore, we omit the second term in~\eqref{equ_sam_grad}.
Consequently, the robustness-enhanced parameter optimization~\eqref{equ_sam} can be solved by
\beq
\label{equ_param_update}
\btheta = \btheta - \alpha \nabla_{\btheta} (\cC(\btheta;\cD_k) + \cR(\btheta; \cK_k))|_{\btheta=\btheta+\bdelta^*},
\eeq
where $\alpha$ denotes the learning rate.

In summary, the complete algorithm for solving the sequential training problem of \name is presented in Algorithm~\ref{alg_overall}.

\input{sections/algorithm/alg_1.tex}

\subsection{Complexity Analysis}\label{s2ec_complex_ana}
We analyze the computational complexity of the training and the core-set selection in the proposed Algorithm~\ref{alg_overall}, as well as the extra memory consumption.
Specifically, we focus on the influence of key parameters, including: sequence length of CSI data $N$, dimension of CSI samples $L_{\rH}$, dimension of latent feature vectors $L$, size of domain datasets $M$, number of classes $C$, number of exemplars per class $E$, clustering-herding ratio $\beta$, and number of training iterations~$I$.

\subsubsection{Training Complexity}
Since the computational cost of the core back-propagation process for training scales linearly with the loss calculation~\cite[Ch. 6.5]{goodfellow2016deep}, we focus on analyzing the complexity of computing the objective function in (P1)$_k$.
For each CSI data, the complexity of predicting its class and computing the loss depends on the architecture of $\bgun(\cdot)$, which mainly comprises FC and MHSA layers.
The complexity of an FC layer is proportional to the product of its input and output dimensions, while an MHSA layer processing a sequence of length $N$ and element dimension $L$ has a complexity of $\cO(N^2L + NL^2)$~\cite{Han21NIPS_Transformer}.
Therefore, the complexity of predicting and calculating the loss for each CSI data is $\cO(NL_{\rH}L + N^2L+NL^2 + NLC)$.
Thus, for the domain dataset $\cD_k$ of $M$ CSI samples and the knowledge core-set of $(k-1)CE$ exemplars from the previous $k-1$ domains, the complexity of calculating the objective function and back-propagation over $I$ training iterations is expressed as:
\beq
\label{equ_complexity_1}
\cO\big( INL\times(N+L_{\rH}+L+C) (M+(k-1)CE) \big).
\eeq

The robustness-enhanced optimization~\eqref{equ_param_update} approximately doubles the computation per iteration due to calculating gradients twice, therefore does not change the overall complexity expression~\eqref{equ_complexity_1}.
Since $L_{\rH}\!>\!L\!\gg\!C$ and $M\gg (k-1) CE$, \eqref{equ_complexity_1} can be simplified to $\cO((N^2L+NL_{\rH}L+NL^2)MI)$, i.e., the training time for solving (P1)$_k$ increases linearly with $M$, $I$, and $L_{\rH}$ and quadratically with $N$ and $L$.

\subsubsection{Core-Set Selection Complexity}
As for the selection of the exemplars in $\cD_k$ after solving (P1)$_k$, our designed algorithm has a low complexity.
It is worth noting that the feature vector for each CSI data can be reused from the last training iteration to form $\cF_{k,c}$, thus incurring no additional cost.
Selecting $\beta E$ exemplars using the K-means clustering method with k-means++ initialization has a complexity of $\cO(\beta EML)$~\cite{Ahmed20Elec_kmeans}, and the complexity of selecting the remaining $(1-\beta)E$ exemplars with the herding method is $\mathcal{O}((1-\beta)EML)$.
Overall, it needs $\cO(CEML)$ computation to obtain the knowledge core-set for $\cD_k$, which is negligible compared to the training complexity.

\subsubsection{Extra Memory Consumption}
We analyze the extra memory required to store the knowledge core-set at the end of the $k$-th period, i.e., $\cK_{k+1}$.
Set $\cK_{k+1}$ consists of $kCE$ exemplars, each comprising a CSI data of size $NL_{\rH}$ and a probability vector of size $C$. 
Consequently, the total memory requirement is $k(NL_{\rH} + C)CE$ floating-point numbers.
Since $CE \ll M$, the memory consumption for storing the knowledge core-set is significantly lower than that required to store the full domain datasets.
 

%% file: sections/analyses/prop_2.tex
\begin{proposition}\emph{(Efficacy of CSI Preprocessing)}
    \label{prop_2_x_data}
    Assume that the UD is in proximity to its user, and that the channel gain of static paths is much larger than that of dynamic paths because of line-of-sight transmission over the static paths.
    For an arbitrary Tx antenna and subcarrier, denote the difference of the pre-processed CSI in~\eqref{equ_x_data} by $\Delta x_{r}[n] = x_{r}[n]-x_{r}[n-1]$. 
    Then, based on~\eqref{equ_csi}, it can be derived that the following relationship holds approximately:
    \beq
    \label{equ_efficacy_x}
    \Delta x_r[n] \propto \Delta h_{\rU, r}[n] + \beta_r \overline{\Delta h_{\rU,1}}[n], 
    \eeq
    where $\Delta h_{U,r}[n]=h_{U,r}[n]-h_{U,r}[n\!-\!1]$ denote the difference of the channel gain at the $r$-th Rx antenna associated with the user's movements, and $\beta_r = h_{\rstatic,r}/\overline{h_{\rstatic,1}}$ with $h_{\rstatic,r}$ being the static channel gain at the $r$-th Rx antenna. 
\end{proposition}
\begin{IEEEproof}
    See Appendix~\ref{proof_prop_2}.
\end{IEEEproof}

%% file: sections/algorithm/alg_1.tex
\begin{figure}[!t]
    \vspace{-0.7em}
    \begin{algorithm}[H]
    \small
    \caption{Robustness-enhanced sequential training with distilled core-set replay for transformer-based discriminator.}
    \label{alg_overall}
    \begin{algorithmic} [1]
        \Require \Copy{R3-4}{\frev{Number of classes $C$, number of exemplars per class $E$, confidence downscaling factor $\eta$, clustering-herding ratio $\beta$, deviation radius $\epsilon$, and learning rate $\alpha$.}}
        \State Initialize $\bgun(\cdot; \btheta)$ based on transformer-based architecture in Sec.~\ref{sec_alg_1} with random parameters, and set $\cK_1=\emptyset$.
        \For{Period $k=1,\dots, K$}
            \State Collect domain dataset $\cD_{k} = \{(\bm H, \hat{\bm p})\}$ for new users. 
            \State With $\cD_{k}$ and $\cK_{k}$, solve the robustness-enhanced optimization~\eqref{equ_sam} by the parameter update in~\eqref{equ_param_update} for $I$ iterations.
            \State Generate feature vector set $\cF_k$ by processing CSI data in $\cD_k$ with $\hat{\bgun}(\cdot;\btheta^*_k)$. 
            \For{Class $c=1,\dots, C$}
                \State Select set ${\cS}_{k,c}$ of $E$ exemplars based on $\cF_{k,c}$ by hybrid clustering and herding method in~\eqref{equ_kmeans} and~\eqref{equ_herding}. 
                \State With $\bgun_{\eta}(\cdot;\btheta^*_k)$, predict probabilities for feature vectors in $\cS_{k,c}$ to obtain $\tilde{\cS}_{k,c}$.
            \EndFor
            \State Update the knowledge core-set by $\cK_{k+1} = \cK_{k} \cup \{\tilde{\cS}_{k,c}\}_{c=1}^C$.
            \State Discard $\cD_k$ to free the memory for the next domain dataset.
        \EndFor
    \end{algorithmic}
\end{algorithm}
\vspace{-1.em}
\end{figure}

%% file: sections/6_simulation.tex
\section{Evaluations}\label{sec_eva}

In this section, we describe the evaluations for the proposed \name framework and the designed algorithm.
We first elaborate on the experimental setup, including details of the adopted multi-user HAR dataset, collected by using a practical wireless ISAC system, and the hyper-parameters in the implementation of the algorithm.
Following that, we present the experimental results, including benchmark comparisons and impact factor analyses.
\begin{figure}[b]
    \vspace{-1em}
	\centering 
	\setlength{\abovecaptionskip}{6pt} 
	\setlength{\belowdisplayskip}{-5pt}
    \subfigure[]
    {
        \centering         
        \includegraphics[width=0.46\linewidth]{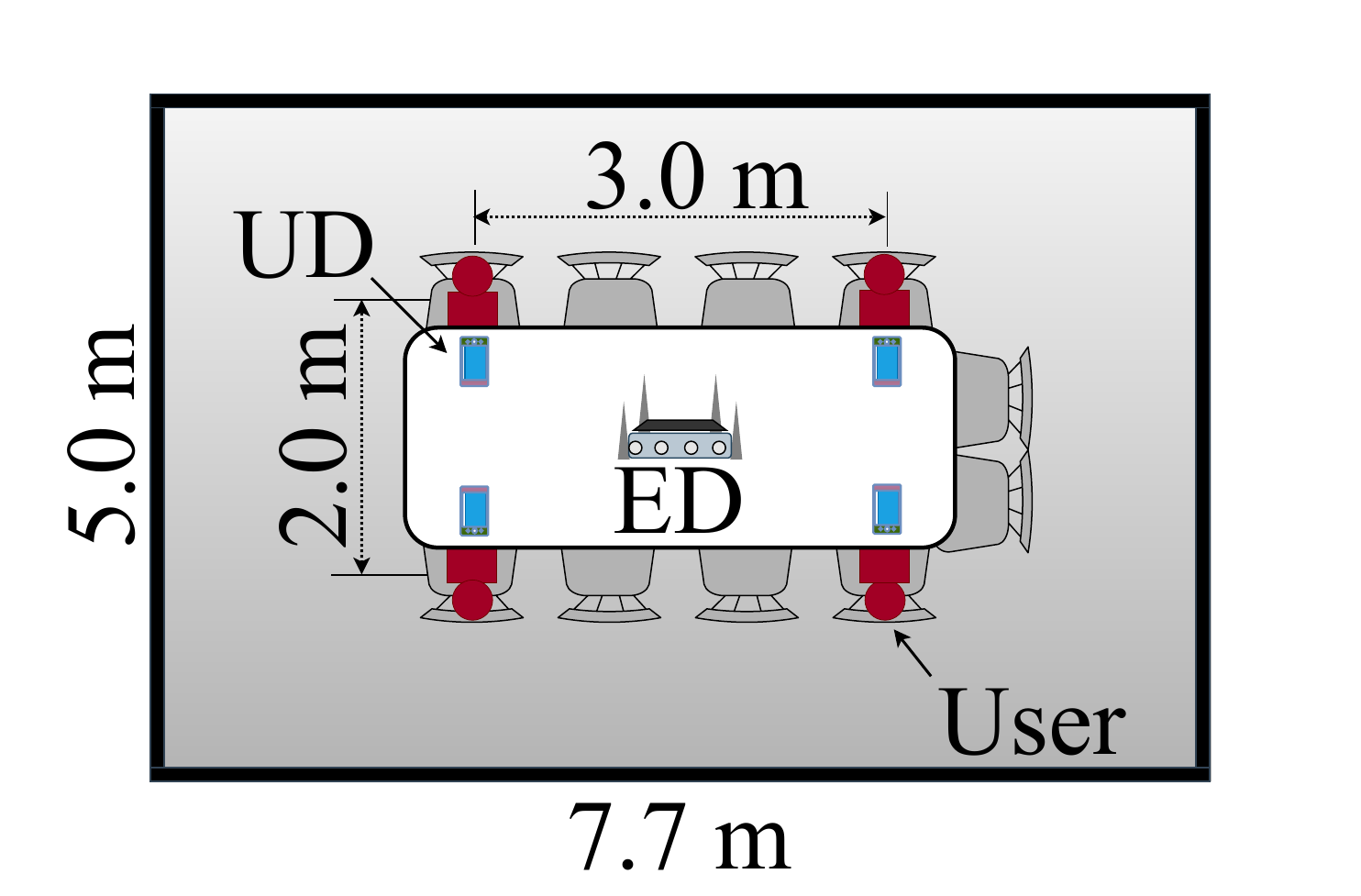}  
        \label{fig:env_meeting} 
    }
        \subfigure[]
    {
        \centering         
        \includegraphics[width=0.46\linewidth]{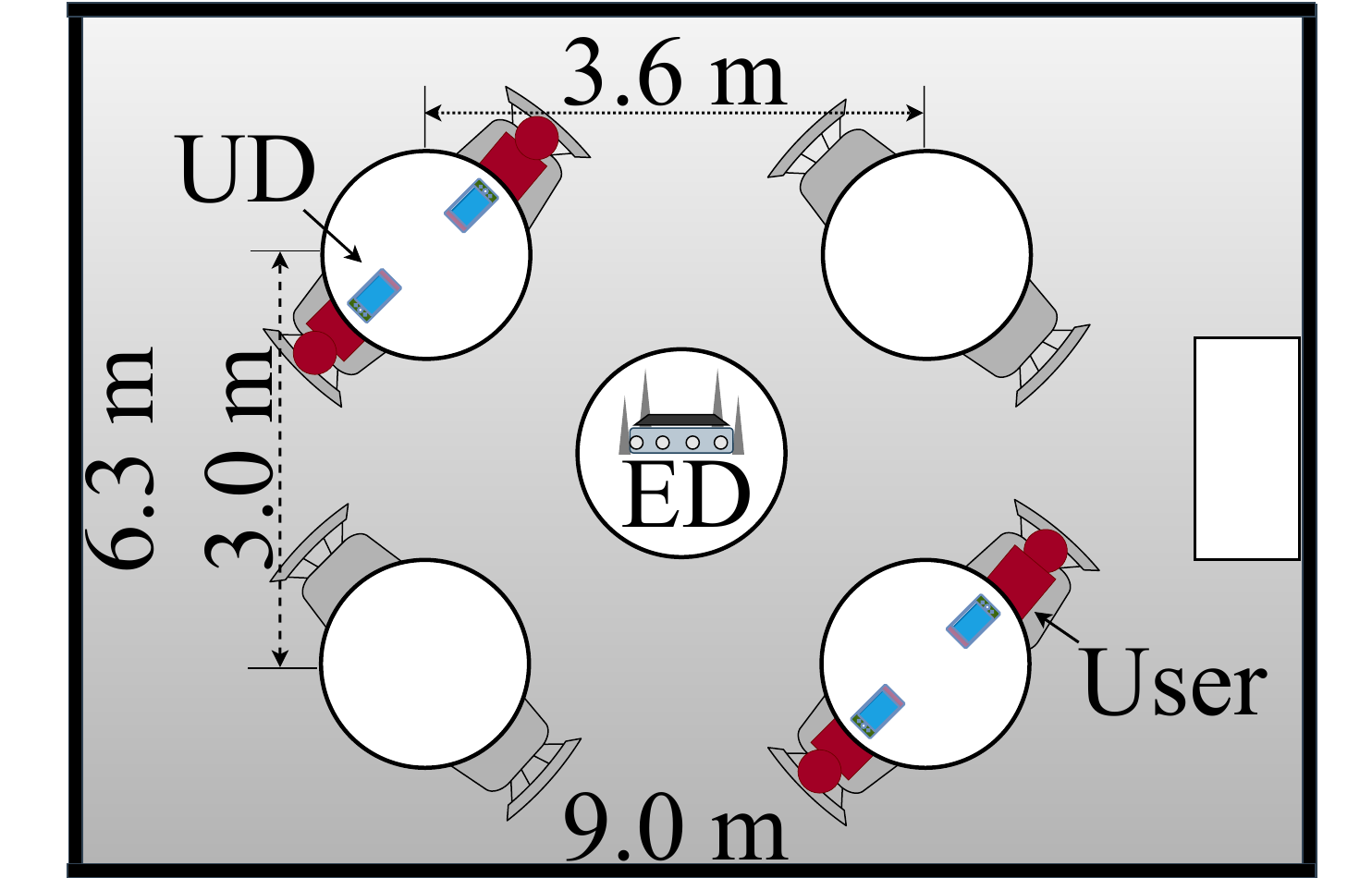}  
        \label{fig:env_class} 
    }
	\vspace{-.7ex}
	\caption{(a) Layouts of the experimental environments in a meeting room and (b) a lecture room.}
	\label{fig:implementOverview}
\end{figure} 
\begin{figure*}[t]
	\Copy{new_fig_4}{
    \centering
    \includegraphics[width=0.9\linewidth]{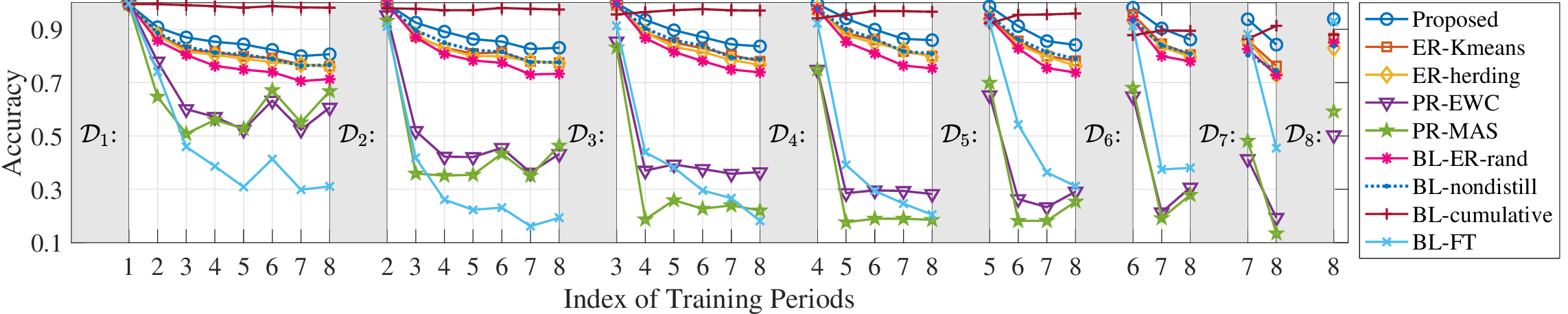}
	\caption{\frev{Comparison between the proposed algorithm and the benchmarks and baselines for the accuracy on each of the eight domain datasets after each training period, averaged over 30 trials.}}}
	\label{fig_overall_compare}
\end{figure*}
\subsection{Experimental Setup}\label{s2ec_exp_setup}

\subsubsection{Dataset Collection}
To evaluate the \name, we collect an HAR dataset by using a practical wireless ISAC system, where multiple users connect to a Wi-Fi access point (AP).
The Wi-Fi network setup by the AP follows IEEE 802.11ac standard~\cite{IEEE_standard}, operating at 5.28\!~GHz and using a bandwidth of $B=40$\!~MHz.
The UDs generate uplink data traffic by engaging in an online Zoom meeting. 
A laptop placed next to the AP emulates the ED and uses PicoScenes~\cite{PicoScenes_IoIJ21} to collect CSI samples from UDs' uplink Quality-of-Service~(QoS) data packets sent to the AP. 
Each CSI sample is a complex array of size $117\times 2$, corresponding to the channel gains for the $117$ subcarriers at two Rx antennas of the AP.

The complete HAR dataset consists of 20 hours of CSI samples for $C=10$ activities performed by $8$ recruited volunteers in two environments, including a meeting room and a lecture room.
The $8$ volunteers consist of $6$ males and $2$ females, and Informed Consent was obtained.
The $10$ activities consist of $4$ body activities: bending, jumping, rotating, and walking; and $6$ gesture activities: push\&pull, sweeping, drawing circle, drawing zigzag, typing-on-phone, and hand-shaking.
In each collection scenario, four volunteers acting as users are present simultaneously to perform the same activities, with each UD positioned around 20\!~cm away from its user, as illustrated in Figs.~\ref{fig:env_meeting} and~\ref{fig:env_class}.
Each activity consists of two seconds of movement followed by one second of rest, leading to CSI sequence duration $T\!=\!3$~\!s.
After segmenting CSI samples into sequences and labeling each sequence according to activities, the complete dataset is constructed.
The domain datasets are obtained by splitting the complete dataset based on individual users.
By default, each domain dataset contains data from a user collected in the two environments.

\subsubsection{Algorithm Hyper-parameters}
For the transformer-based discriminator in Sec.~\ref{sec_alg_1}, we implement it with the following hyper-parameters:
i) in the CSI pre-processor, the length of temporal feature vector is $L_{\rT} = 16$; 
ii) in the MLP encoder, there are two FC layers, sequentially encoding each input vector to feature vectors of sizes first $128$ and then $64$;
iii) in the transformer encoder, there are two MHSA and FF layers, with number of heads $A=8$ and feature vector length $L=64$;
iv) in the probability predictor, the first $64$-dim feature vector in each sequence is mapped to $C=10$ logits and then converted to probabilities by a default softmax function.

Before training on the first domain dataset, the neural parameters are initialized using the Xavier uniform distribution as in~\cite{Glorot10ICAIS_Understanding}.
By default, during the training of each domain dataset, the learning rate is $\alpha=10^{-3}$, and the number of training iterations is $I=500$. 
In addition, the dropout rate for feature elements is set to $0.1$ to improve generalizability.
The deviation radius in the robustness-enhanced optimization is set to $\epsilon=0.03$.
In each domain dataset, the number of exemplars to select for each class is $E=10$, and the clustering-herding ratio during exemplar selection is $\beta=0.9$.
The confidence downscaling factor is set to $\eta=2$.

\subsection{Experimental Results}\label{s2ec_exp_res}
Below, we present the experimental results on \name from two perspectives, including the overall benchmark comparison and impact factor analysis.
Each experiment is conducted with 30 trials using different random seeds to account for randomness in stochastic training and clustering.
\Copy{R3-5}{\frev{By default, we use the complete domain dataset in each training period to train the neural model and consistently measure the accuracy on that dataset in subsequent training periods, which facilitates a low-variance evaluation of knowledge retention.}}

\subsubsection{Overall comparison}\label{s3ec_overall_compare}

Firstly, we split the CSI data for the eight users into eight domain datasets, resulting in an average domain dataset size of $M=3013$.
We train the neural model sequentially on the eight domain datasets, using the \name framework with the proposed algorithm, and compare its performance against four state-of-the-art benchmarks from the two most common categories of continual learning methods: ER-based and PR-based methods.
\begin{itemize}[leftmargin=*] 
\item \textbf{ER-Kmeans}: Core-sets of exemplars are selected by the K-means clustering method~\cite{Nguyen18ICLR_VCL}, and their empirical CE loss is included in the training as the regularization function.
\item \textbf{ER-herding}: Core-sets of exemplars are selected by the herding method, and their empirical CE loss is included in the training as the regularization function~\cite{Rebuffi17CVPR_iCaRL}.
\item \textbf{PR-EWC}: The regularization function in~\eqref{equ_R_G} is used for knowledge retention, where the variance vector is derived by the elastic weight consolidation~(EWC) method~\cite{Kirkpatrick17PNAS_Overcome}. 
\item \textbf{PR-MAS}: The regularization function in~\eqref{equ_R_G} is used for knowledge retention, where the variance vector is derived by the memory-aware synapses~(MAS) method~\cite{Aljundi18ECCV_Memory}.
\end{itemize}

Fig.~\ref{fig_overall_compare} shows the comparison results in terms of the classification accuracy.
\Copy{R1-9}{\frev{In Fig.~\ref{fig_overall_compare}, the x-axis is divided into eight regions, showing the accuracy of the neural model on $\cD_1$ to $\cD_8$, respectively. 
In the $k$-th region~($k=1,...,8$), the x-axis range spans the training periods from $k$ to $8$, representing the model's accuracy on $\cD_k$ after it is firstly trained on $\cD_k$ in the $k$-th period and subsequently trained on each $\cD_{k'}$ ($k<k'\leq 8$) in later periods.}}
\Copy{R1-7}{\frev{Fig.~\ref{fig_overall_compare} shows that the proposed algorithm outperforms the four bulleted continual learning benchmarks, resulting in an average accuracy of $0.853$, which surpasses the second-best (ER-Kmeans, $0.787$) by a noticeable $8\%$.}}
Besides, for the proposed algorithm, the average accuracy loss for $\cD_1,\dots,\cD_7$ due to forgetting is $0.144$, which is $21\%$ less than $0.183$ of the second-best.
Moreover, our evaluation confirms that, for CSI processing, the ER-based approaches also significantly outperform the PR-based ones, $0.806$ vs. $0.356$ in average.

Furthermore, we compare four baselines~(BL) in Fig.~\ref{fig_overall_compare}.
\Copy{R2-1}{\frev{We first evaluate the ER method with random core-set selection, named BL-ER-rand, showing its average results given 30 random seeds. It can be observed that randomly selected samples result in less effective exemplars for knowledge retention compared with those selected by herding, K-means clustering, and the proposed method.}}
\Copy{R2-4}{\frev{In addition, we evaluate the proposed method with non-distilled core-set, where the exemplars have hard one-hot labels, referred to as the BL-nondistill baseline. 
As shown in Fig.~\ref{fig_overall_compare}, BL-nondistill exhibits a decline in accuracy compared with the proposed method, highlighting the importance of using distilled core-set for knowledge retention.}}
The other two baselines include the sequential fine-tuning baseline, named BL-FT, and the cumulative baseline, named BL-cumulative.
\Copy{R1-8}{\frev{In BL-FT, the model is firstly trained on $\cD_1$ with a learning rate $\alpha$ and then fine-tuned on each subsequent domain dataset with a reduced learning rate of $0.1\alpha$, without any continual learning schemes.}}
In BL-cumulative, all the previous domain datasets are fully stored and re-trained in each period.
As expected, BL-FT results in catastrophic forgetting: the accuracy for previous domains drops rapidly in subsequent training. 
Nevertheless, its outperforms PR-based methods for $\cD_5$ to $\cD_8$, as parameter regularization for previous domains severely undermines the ability of the neural model to learn new domain datasets.

Another interesting observation is that though BL-cumulative completely preserve the accuracy for previous domains, the re-training over previous domain datasets hinders it adapting to the current domain.
Comparing the proposed algorithm with the BL-FT for $\cD_1$ to $\cD_7$ after the eighth training, the proposed algorithm alleviates the catastrophic forgetting of BL-FT by $79\%$.
In addition, comparing the proposed algorithm with the BL-cumulative, it can be calculated that the proposed algorithm achieves $89\%$ of the accuracy for $\cD_1$ to $\cD_7$ with only $3\%$ memory consumption, storing $100$ instead of average $3013$ CSI data for each domain dataset.

\begin{figure}[t]
	\centering 
	\setlength{\abovecaptionskip}{6pt} 
	\setlength{\belowdisplayskip}{-5pt}
	\subfigure[]
	{
		\centering         
		\includegraphics[width=0.47\linewidth]{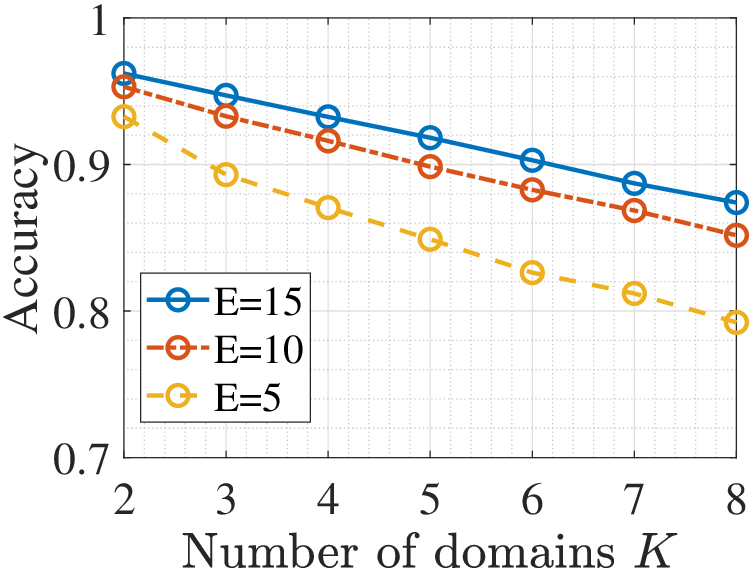}  
		\label{fig_num_domain} 
	}
	\hspace{-10pt}
	\subfigure[]
	{
		\centering
		\includegraphics[width=0.46\linewidth]{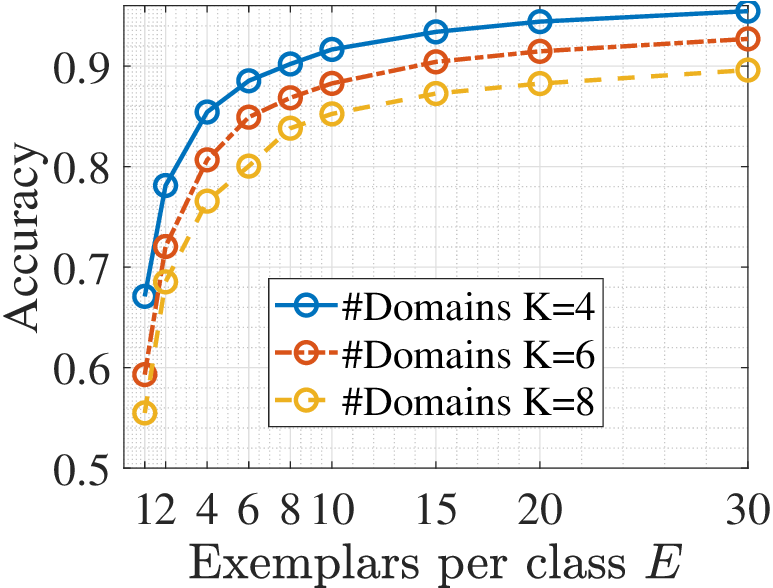}
		\label{fig_buff_size} 
		\captionsetup{justification=centering}
	}
	\vspace{-.7ex}
	\caption{Impact of (a) number of domains $K$ and (b) number of exemplars per class on the average accuracy across the eight domains and 30 trials.}
	\vspace{-.8ex}
\end{figure} 
\subsubsection{Impact Factor Analyses}
We then investigate the impact of six factors on the proposed algorithm and the \name framework, including 
a) number of domains $K$,
b) number of exemplars $E$,
c) number of training iterations $I$, 
d) learning rate $\alpha$,
e) clustering-herding ratio $\beta$,
and f) deviation radius $\epsilon$.

Fig.~\ref{fig_num_domain} shows that as the number of domains $K$ increases, the average accuracy for the domain datasets after training decreases linearly.
Nevertheless, with more exemplars allowed, the decrease rate can be reduced.
On the other hand, Fig.~\ref{fig_buff_size} reveals a promising sign that using a few exemplars ($E=10$) achieves comparable average accuracy to that of a much larger number ($E=30$), which is consistent across different values of $K$.
These observations indicate that the proposed algorithm has the potential to help the EI handle large numbers of domains with a small number of exemplars.

\begin{figure}[b]
	\centering 
	\setlength{\abovecaptionskip}{6pt} 
	\setlength{\belowdisplayskip}{-5pt}
	\subfigure[] 
		{
		\centering
		\includegraphics[width=0.46\linewidth]{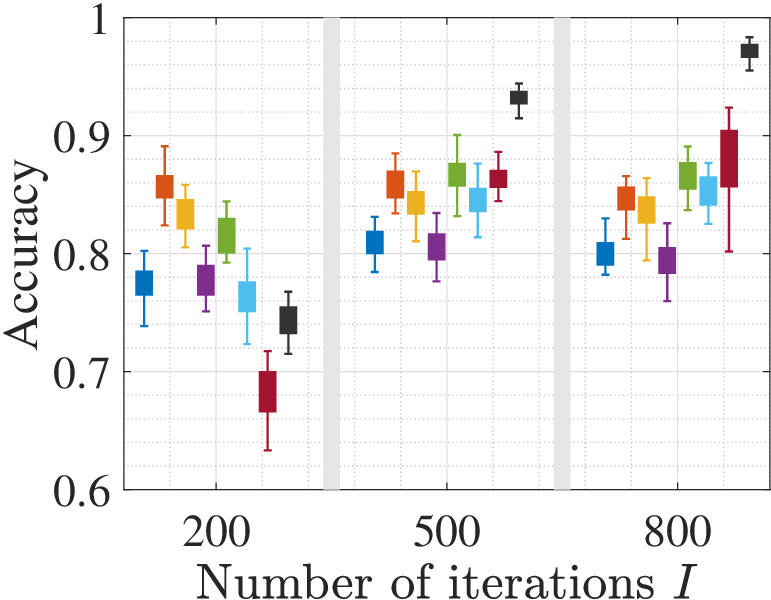}
		\label{fig_iter_compare} 
		\captionsetup{justification=centering}
		}
		\hspace{-8pt}
        \subfigure[]
		{
		\centering         
		\includegraphics[width=0.46\linewidth]{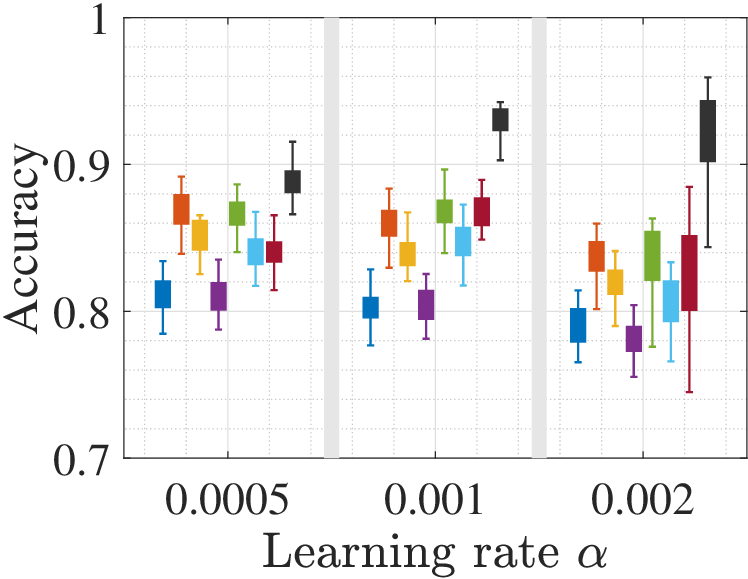}  
		\label{fig_lr_compare} 
		\captionsetup{justification=centering}
        }
	\vspace{-.7ex}
	\caption{Impact of (a) number of training iterations $I$ and (b) learning rate $\alpha$ on the average accuracy across the eight domains. Box plots indicate the distribution over the 30 trials.}
	\vspace{-.8ex}
\end{figure} 

\begin{figure}[t]
    \vspace{-.5em}
	\centering 
	\setlength{\abovecaptionskip}{6pt} 
	\setlength{\belowdisplayskip}{-5pt}
	\subfigure[]
	{
		\centering         
		\includegraphics[width=0.47\linewidth]{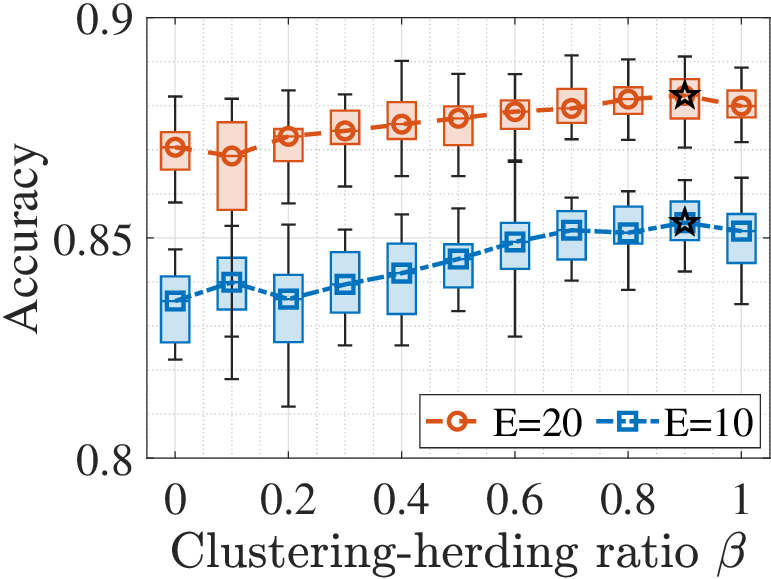}  
		\label{fig_kr_ratio_influen} 
		\captionsetup{justification=centering}
	}
	\hspace{-14pt}
	\subfigure[]
	{
		\centering
		\includegraphics[width=0.47\linewidth]{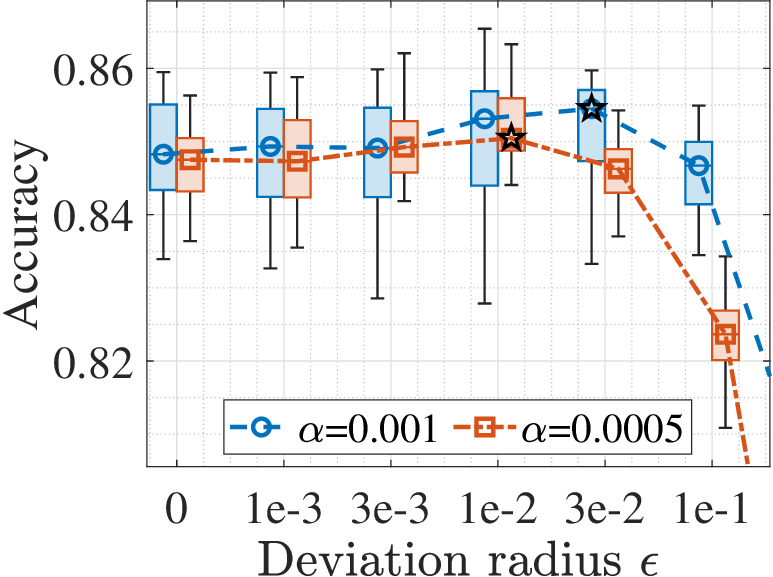}
		\label{fig_eps_influen} 
		\captionsetup{justification=centering}
	}
	\vspace{-.7ex}
	\caption{Impact of (a) maximum deviation distance $\epsilon$ and (b) on accuracy after training. Box plots indicate the distribution over the 30 trials, curves connect median values, and stars represent the highest values.}
	\vspace{-.8ex}
\end{figure} 

Figs.~\ref{fig_iter_compare} and~\ref{fig_lr_compare} demonstrate how the training depth and intensity, represented by the number of iteration $I$ and learning rate $\alpha$, respectively, impact the accuracy for each domain after the eight periods of training.
Specifically, the eight box plots, from left to right, in each case represent the accuracy distributions for $\cD_1$ to $\cD_8$ over the 30 trials.
Fig.~\ref{fig_iter_compare} shows that increasing the number of iterations $I$ from $200$ to $500$ improves the accuracy for all the eight domain datasets.
However, when $I$ is further increased to $800$, the accuracy for earlier domain datasets ($\cD_1$ to $\cD_4$) tends to decrease.
Notably, while a moderate increase in learning depth improves the learning for the current domain datasets, an excessive depth may cause more forgetting of previously learned knowledge.
Fig.~\ref{fig_lr_compare} compares the accuracy distributions of the eight domains for different learning rate $\alpha$.
Between $\alpha=0.0005$ and $\alpha=0.001$, a smaller learning rate improves accuracy on the earlier domain datasets, i.e., $\cD_1$ to $\cD_4$, but decreases it for the later ones, i.e., $\cD_5$ to $\cD_8$.
This is probably because a small learning rate prevents the parameters from deviating from the training results of previous domains, thereby aiding in the preservation of previous knowledge. 
Moreover, using a large learning rate, e.g., $\alpha=0.002$ results in general accuracy drop and training instability.
Nevertheless, it is worth noting that the proposed algorithm remains relatively robust to variations in training depth and intensity over $I\in[500,800]$ and $\alpha\in[0.0005,0.001]$.

Figs.~\ref{fig_kr_ratio_influen} and~\ref{fig_eps_influen} show how the two key hyper-parameters of the proposed algorithm, i.e., the clustering-herding ratio $\beta$ in core-set selection and the deviation radius $\epsilon$ in the robustness-enhanced optimization, impact the average accuracy.
Fig.~\ref{fig_kr_ratio_influen} demonstrates that the accuracy increases as more exemplars are selected by clustering, while drop when $\beta=1$.
Therefore, the best strategy for core-set selection is to select most of them by clustering while using a few to ensure the closeness between the center of core-set and that of the full domain dataset in the latent feature space.
Fig.~\ref{fig_eps_influen} confirms that by optimizing the maximum loss within a small deviation radius $\epsilon\in(0,0.03)$, the average accuracy can be improved.
It is worth noting that when $\epsilon$ gets larger, e.g., $\epsilon\geq 0.03$, the resulting accuracy drops rapidly since the first-order approximation in~\eqref{equ_sam_delta} becomes highly inaccurate for large $\epsilon$.
Furthermore, it can also be observed that the proper deviation radius for a smaller learning rate should also be smaller.

%% file: sections/7_conclusion.tex
\section{Conclusion}
\label{sec_conclu}

This paper has proposed the \name framework to enable cross-domain continual learning for EI in wireless ISAC networks to handle discriminative tasks.
First, we have modeled the CSI data and analyzed its domain specificity.
Given the limited memory and computational resources of EDs, which cannot support joint re-training on all domain datasets, \name reformulates cross-domain continual learning as a sequential training problem.
We have designed a novel efficient algorithm to solve this problem, integrating the transformer-based discriminator for CSI sequences, knowledge retention through replaying distilled core-sets selected by hybrid clustering and herding, and robustness-enhanced optimization to mitigate accuracy decline due to parameter deviations.
On a real-world ISAC HAR dataset, we have validated that \name mitigates the accuracy decline due to catastrophic forgetting by $79\%$, achieving $89\%$ of the cross-domain training performance while using only $3\%$ of memory of the cumulative training.

%% file: sections/appendices/app_prop.tex
\section{Proof of Proposition~\ref{prop_2_x_data}}
\label{proof_prop_2}

Substitute~\eqref{equ_csi} into~\eqref{equ_x_data}, and omit the gains for the environmental dynamic paths and noise, as they are dominated by the dynamic channel gains for the user's motions in the proximity to the UD.
Then, $x_{r}[n]$ can be expressed as
\begin{align}
\label{equ_exp_xr_1}
x_{r}[n]  &\approx\norm( h_{\rstatic,r}\overline{h_{\rstatic,1}} + h_{\rU,r}[n]\overline{h_{\rstatic,1}} + \overline{h_{\rU,1}}[n]h_{\rstatic,r})\\
\label{equ_exp_xr_2}
& = \frac{h_{\rU,r}[n]+\beta_r \overline{h_{\rU,1}[n]}}{h_{\rU,r}[n^*]+\beta_r \overline{h_{\rU,1}}[n^*]}\\
\label{equ_exp_xr_3}
& \propto h_{\rU,r}[n]+\beta_r \overline{h_{\rU,1}}[n].
\end{align}
Here, $n^* = \argmax_{n'} |h_{\rU,r}[n']+\beta_r \overline{h_{\rU,1}}[n']|$
We note that in~\eqref{equ_exp_xr_1}, the term $h_{\rU,r}[n]\overline{h_{\rU,1}}[n]$ is also omitted based on the assumption that the gain of the static paths is much larger than that of dynamic ones.
Substituting~\eqref{equ_exp_xr_2} into the expression of $\Delta x_{r}[n]$ proves Proposition~\ref{prop_2_x_data}. \hfill $\qed$